**Title**

Modelling $N_2O$ dynamics of activated sludge biomass under nitrifying and denitrifying conditions: pathway contributions and uncertainty analysis.

**Author list**

Carlos Domingo-Félez, Barth F. Smets*

Department of Environmental Engineering, Technical University of Denmark, Miljøvej 115, 2800 Kgs. Lyngby, Denmark

* Corresponding author:

Barth F. Smets, Phone: +45 4525 1600, Fax: +45 4593 2850, E-mail: bfsm@env.dtu.dk

**Abstract**

Nitrous oxide ($N_2O$) is a potent greenhouse gas emitted during biological wastewater treatment. A pseudo-mechanistic model describing three biological pathways for nitric oxide (NO) and $N_2O$ production was calibrated for mixed culture biomass from an activated sludge process using laboratory-scale experiments.

The model (NDHA) comprehensively describes $N_2O$ producing pathways by both autotrophic ammonium oxidizing bacteria and heterotrophic bacteria. Extant respirometric assays and anaerobic batch experiments were designed to calibrate endogenous and exogenous processes (heterotrophic denitrification and autotrophic ammonium/nitrite oxidation) together with the associated net $N_2O$ production. Ten parameters describing heterotrophic processes and seven for autotrophic processes were accurately estimated (variance/mean < 25%). The model predicted NO and $N_2O$ dynamics at varying dissolved oxygen, ammonium and nitrite levels and was validated against an independent set of experiments with the same biomass.

Aerobic ammonium oxidation experiments at two oxygen levels used for model evaluation (2 and 0.5 mg/L) indicated that both the nitrifier denitrification (42, 64%) and heterotrophic denitrification (7, 17%) pathways increased and dominated $N_2O$ production at high nitrite and low oxygen concentrations; while the nitrifier nitrification pathway showed the largest contribution at high dissolved oxygen levels (51, 19%). The uncertainty of the biological parameter estimates was propagated to $N_2O$ model outputs via Monte Carlo simulations as 95% confidence intervals. The accuracy of the estimated parameters resulted in a low uncertainty of the $N_2O$ emission factors (4.6 ± 0.6% and 1.2 ± 0.1%).

**Highlights**

- A model describing three biological $N_2O$ and NO production pathways is calibrated with an activated sludge biomass.
- In respirometric assays $N_2O$ and NO production increased during $NH_4^+$ oxidation under low DO and the presence of $NO_2^-$.
- Aerobic NH4 oxidation-driven N2O production increased at low DO and in the presence of $NO_2^-$.
- The uncertainty of biological parameter estimates was evaluated for $N_2O$ model predictions.

## 1. Introduction

Nitrous oxide ($N_2O$) is a greenhouse gas emitted at wastewater treatment plants. During biological nitrogen removal $N_2O$ is mainly emitted from aerated zones or during aerated periods due to physical stripping (Ahn et al., 2010; Lim and Kim, 2014). $N_2O$ is produced biologically by two microbial guilds: autotrophic ammonia oxidizing bacteria (AOB) and heterotrophic denitrifying bacteria (HB). AOB aerobically oxidize ammonium ($NH_4^+$) into nitrite ($NO_2^-$) via hydroxylamine ($NH_2OH$). Pure culture studies of model AOBs have shown that a side- reaction of hydroxylamine oxidation to nitrite (HAO-mediated) can produce nitric oxide (NO) and $N_2O$ regardless of dissolved oxygen (DO) levels (Caranto and Lancaster, 2017; Kozlowski et al., 2016). Also, direct $N_2O$ production from hydroxylamine oxidation catalysed by the enzyme cytochrome (cyt) P460 was recently documented (Caranto et al., 2016). At low DO concentrations AOBs use nitrite as terminal electron acceptor producing NO, further reduced to $N_2O$ (de Bruijn et al., 1995; Kester, 1997). Heterotrophic denitrification of nitrate ($NO_3^-$) to dinitrogen gas ($N_2$) is a 4-step process via $NO_2^-$, NO and $N_2O$ (Berks et al., 1995).

In nitrogen removing systems $N_2O$ production is associated to high $NH_4^+$ and $NO_2^-$, and low DO concentrations (Kampschreur et al., 2009). The $NH_4^+$ load and influent $NH_4^+$ concentration have been correlated to $N_2O$ emissions from aerobic zones operating at high DO concentrations (Lotito et al., 2012; Ni et al., 2013). At low DO $NH_4^+$ is oxidized at a lower rate but a higher fraction is converted to $N_2O$ (Burgess et al., 2002; Li and Wu, 2014). $NO_2^-$ accumulation also lead to higher $N_2O$ emissions in N-removing systems (Kampschreur et al., 2008; Y. Wang et al., 2016). However, as the direct precursor of $N_2O$ in most of the biological pathways, NO has shown the highest correlations with $N_2O$ (Domingo-Félez et al., 2014; Kampschreur et al., 2008; Y. Wang et al., 2016). While NO emissions represent a very small fraction of the N removed it can accumulate to similar concentrations as $N_2O$ (Schulthess et al., 1995; Wunderlin et al., 2012).

In systems with low nutrient dynamics (e.g. CSTR) batch experiments are more informative than continuous operation and preferred over continuous systems for parameter estimation. Also, lab-scale experiments allow more controlled environments compared to full-scale systems. Among lab-scale experimental designs, accurate kinetic parameter values for heterotrophic aerobic growth and nitrification have been obtained via respirometric techniques (Chandran and Smets, 2000; Petersen et al., 2001). Hence, the use of respirometric experiments to quantify $N_2O$ dynamics from mixed liquor biomass will be assessed.

AOB and HB have been suggested as equal contributors to $N_2O$ production during traditional nitrification and denitrification processes (Hu et al., 2011; Jia et al., 2013; Lotito et al., 2012; Tallec et al., 2006). However, in-situ quantification via nitrogen and oxygen isotopic signatures is rare in full-scale systems compared to lab-scale (Toyoda et al., 2011; Wunderlin et al., 2013). To gain insights on the individual process contributions $N_2O$ emissions are correlated to operational parameters (e.g. $NH_4^+$ oxidation rate) or by multiple linear regression to operational factors (Leix et al., 2017).

The increasing metabolic understanding of nitrogen removal can be described with mathematical equations and has been successfully used to predict the fate of nitrogenous species in wastewater treatment operations (Henze et al., 2000). Mechanistic models for N-removal have been extended to include $N_2O$ production processes which allow the prediction of individual pathway contributions. Three biological $N_2O$ production pathways are considered to co-occur: the nitrifier nitrification (NN), the nitrifier denitrification (ND) and the heterotrophic denitrification (HD) pathway. Several models have been published, with varying number of processes and variables considered, and different mathematical description of the process rates. Models considering the three pathways have better predictive capabilities than two pathway models (Spérandio et al., 2016). The NDHA model considers the three biological pathways and abiotic contribution and was previously calibrated for an AOB-enriched biomass (Domingo-Félez et al., 2017a). Here, we aim at calibrating the NDHA model for a mixed culture biomass from a full-scale wastewater treatment plant via respirometric assays.

A key characteristic of current $N_2O$ models is the variability of kinetic parameter values (e.g. $K_{HB.NO}$ = 0.00015 – 0.05mgN/L) (Hiatt and Grady, 2008; Pan et al., 2013), which has been associated to different structure simplifications or microbial population switches (Spérandio et al., 2016). The high variability affects the accuracy of $N_2O$ predictions as newly estimated parameters depend on fixed parameter values. However, only best-fit simulations and not the uncertainty of $N_2O$ models on full-scale systems has been reported, which will benefit the design of $N_2O$ mitigation strategies.

**Objectives**

- Quantify $N_2O$ dynamics from AS biomass via extant respirometric assays.
- Calibrate the NDHA model to describe N-removing processes and $N_2O$ production of the AS biomass and assess the accuracy of estimated parameters.
- Evaluate the predictive ability of the calibrated model against a different AS biomass dataset.
- Quantify the uncertainty of $N_2O$ emissions during aerobic $NH_4^+$ removal at two DO concentrations.

Mixed Liquor WWTP → Lab-scale Respirometry → Model Calibration / Validation → $N_2O$ Pathway contribution?

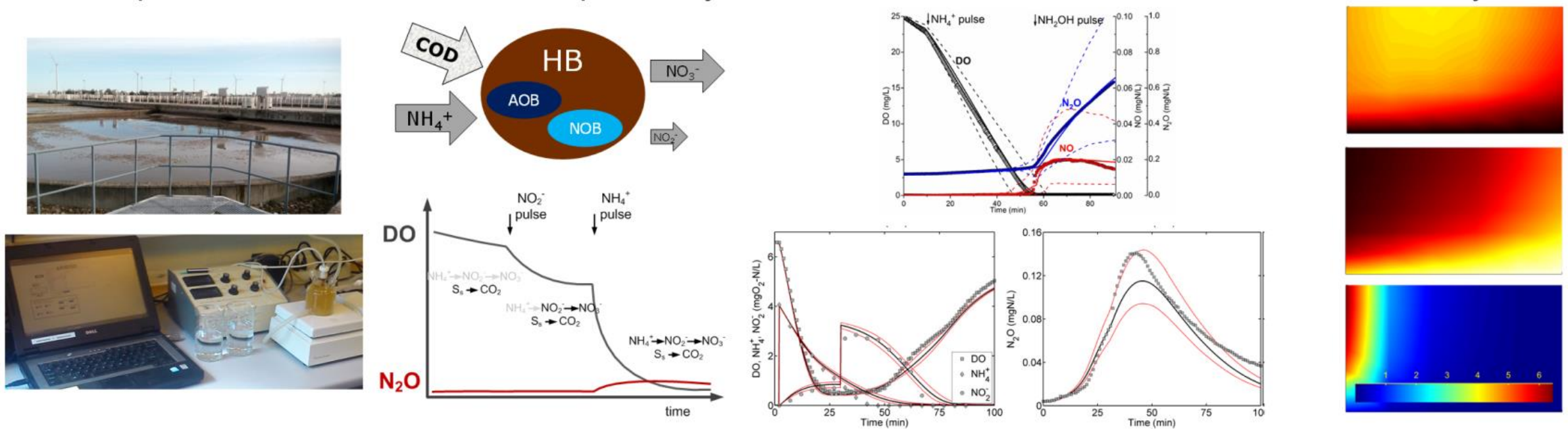

## 2. Materials and Methods

### 2.1. Extant respirometric assays

Mixed liquor from a full-scale phase-isolated activated sludge (AS) wastewater treatment plant (700,000 PE Lynetten, Copenhagen, Denmark) was sampled and aerated overnight; it is here referred to as the AS biomass. To prevent significant changes in biomass composition biomass was discarded, and new samples taken, after two days of experimentation (Li and Ju, 2002). Biomass samples were harvested by centrifugation at 4700 rpm for 3 min, washed and resuspended in nitrogen-free mineral medium (modified after Graaf et al., 1996) three times to eliminate any soluble substrates. Assays were performed in parallel at 25°C in two 400-mL jacketed glass vessels completely filled with biomass and sealed with the insertion of a Clark-type polarographic DO electrode (YSI Model 5331, Yellow Springs, OH). Biomass samples were saturated with air, pure oxygen or dinitrogen gas prior to the initiation of the assays. A decrease in the DO level in the vessel due to substrate oxidation was measured by the DO probe and continuously acquired (0.2 Hz) by a personal computer interfaced to a DO monitor (YSI Model 5300, Yellow Springs, OH) by a multi-channel data acquisition device (LabPC+, National Instruments, Austin, TX). pH was also monitored (WTW GmbH, Weilheim, Germany) and liquid NO and $N_2O$ concentrations were measured with Clark-type microsensors (NO-500, N2O-R, Unisense A/S, Aarhus, Denmark). Stock solutions for all the reagents were prepared from high-purity chemicals for $NH_4HCO_3$, $NH_2OH{\cdot}HCl$, $NaNO_2$, HCOONa, $C_2H_3NaO_2$, $C_3H_5NaO_2$ (Sigma Aldrich) and by initially sparging ≥ 99.998%, 99.5% gas $N_2O$ and $O_2$ (Sigma-Aldrich, AGA). Photometric test kits were used to analyse N species (1.14752, 1.09713, 1.14776, Merck KGaA, Darmstadt, Germany). Biomass content (MLSS, MLVSS) was measured in duplicates according to APHA (APHA et al., 1999) and estimates of nitrifying fraction were made from stoichiometric principles (Domingo-Félez et al., 2017b).

### 2.2. Experimental Design

The experiments were designed to obtain informative data on $N_2O$ dynamics from mixed liquor biomass. Respirometric approaches were taken (on-line, high-rate $O_2$ and $N_2O$ measurements) as they are more suited for accurate parameter estimation in comparison to substrate depletion experiments (Chandran et al., 2008). The kinetics of the oxidation of the primary N-species ($NH_4^+$ and $NO_2^-$) were individually and step-wise measured via extant respirometry at varying DO concentrations, while simultaneously measuring $N_2O$. Batch experiments were conducted to measure the heterotrophic and abiotic contributions to total $N_2O$ production (**Table 1**). During anoxic experiments processes were measured both in absence and presence of supplied organic carbon (mixture of formate + acetate + propionate) under $NO_x^-$ excess. With no organic carbon supply $NO_x^-$ reduction was assumed supported by hydrolysis products originated from biomass decay.

A scenario (e.g. (Scenario C)) was defined as a group of experiments with the same primary N-species added by pulses (**Table 1**). At least duplicate experiments were performed for each N-species. The oxygen consumption rate was the additive effect of several independent oxygen consumption processes, potentially including endogenous respiration, $NO_2^-$ respiration, and $NH_4^+$-respiration (**S-I**). By sequentially following the respirometric response from more to less oxidized N-species (i.e., first $NO_2^-$, then $NH_4^+$) the identifiability of nitrification kinetic parameters increases (Brouwer et al., 1998).

**Table 1** – Experimental design for respirometric assays (Scenarios A-C) and for model validation (Scenarios D-E).

| Scenario | Oxygen level | Species added | Species monitored | Targeted Processes |
|---|---|---|---|---|
| (A) | Anoxic | COD+$NO_3^-$ / COD+$NO_2^-$ / COD+$N_2O$ | $NO_3^-$, $NO_2^-$, $N_2O$, $NH_4^+$ | Heterotrophic denitrification, hydrolysis |
| | Non-aerated: from excess DO (air-sat) into anoxia dynamically | COD | DO | Biomass decay, hydrolysis |
| (B) | Anoxic | $NO_3^-$ / $NO_2^-$ | $N_2O$, NO | HB-driven NO/$N_2O$ dynamics |
| (C) | Non-aerated: from excess DO ($O_2$-sat) into anoxia dynamically | $NH_4^+$ / $NH_2OH$ / $NO_2^-$ | DO, $N_2O$, NO | $NH_4^+$, $NO_2^-$ removal<br>AOB/HB-driven NO/$N_2O$ dynamics |
| (D) | Constant aeration (high and low DO) | $NH_4^+$ | DO, $N_2O$, $NH_4^+$, $NO_2^-$ | $NH_4^+$, $NO_2^-$ removal, $N_2O$ dynamics |
| (E) | Constant aeration (high and low DO) | $NH_4^+$ / $NO_2^-$ / $NO_3^-$ | DO, $N_2O$, $NH_4^+$, $NO_2^-$ | $NH_4^+$, $NO_2^-$ removal, $N_2O$ dynamics |

### 2.3. Dataset for model evaluation

Separate batch experiments were executed in a 3-L lab-scale reactor with mixed liquor biomass from the same WWTP (Lynetten, Copenhagen, Denmark): (D) Four batch tests received sequential increasing $NH_4^+$ pulses while subject to constant aeration (2.3, 3.6, 4.7, 5.5 mgN/L); (E) three batch tests received a singular $NH_4^+$ pulse (4 mg$NH_4^+$-N/L) followed by addition of either $NO_2^-$ or $NO_3^-$ when reaching low DO concentrations. The dataset comprises online DO, pH and $N_2O$ liquid measurements together with grab samples for $NH_4^+$ and $NO_2^-$. Details of these experiments have been reported before (Domingo-Félez et al., 2017b).

### 2.4. Model description

The NDHA model was proposed to describe NO/$N_2O$ dynamics under a variety of conditions for biomass containing both autotrophic (ammonium and nitrite oxidizing bacteria) and heterotrophic fractions (Domingo-Félez and Smets, 2016). Shortly, it considers $N_2O$ formation from two autotrophic and one heterotrophic biological pathway, plus abiotic $N_2O$ formation based on recent findings (Soler-Jofra et al., 2016). Unlike any other model, NDHA qualitatively captures NO and $N_2O$ profiles observed at both high and low DO. The observed $N_2O$ production during $NH_4^+$ oxidation at high DO is described by the Nitrifier Nitrification pathway (NN) (Ni et al., 2013). The Autotrophic Denitrification (ND) pathway captures the increasing $N_2O$ production at low DO and in the presence of $NO_2^-$. The higher $N_2O$ yield ($N_2O$ produced / N-species consumed) from $NH_2OH$ oxidation, as observed for pure and mixed AOB cultures, is also captured by the model as $NH_2OH$ is the electron donor in both NN and ND pathways (Caranto et al., 2016; Kozlowski et al., 2016; Terada et al., 2017). A 4-step heterotrophic denitrification model was considered based on earlier reports (Hiatt and Grady, 2008). Individual process rates and inhibition/substrate coefficients were used as suggested for systems with low substrate accumulation. Here we aim to calibrate the NDHA model for AS biomass (**S-I**).

### 2.5. Sensitivity analysis and uncertainty evaluation

A global sensitivity analysis (GSA) was performed to identify the most determinant parameters for model outputs via Monte Carlo simulations. Uncertainty from model parameters is propagated as 10-25-50% uniform variations from their default value to model outputs (Sin et al., 2009) (**S-II**). Latin hypercube sampling (LHS) was used to cover the parameter space and the Standardized Regression Coefficient method was used to calculate the sensitivity measure $\beta_i$, which indicates the effect of the parameter on the corresponding model output (convergence found for 1200 samples, $\beta_2$ threshold > 0.7) (Campolongo and Saltelli, 1997). The duration of every experiment was discretized in 400 time steps, and the GSA run at each point, obtaining a dynamic profile of global sensitivity metrics.

The effect of the estimated uncertainty in kinetic parameters on the model output was evaluated by Monte Carlo simulations. Parameter values were sampled via LHS (n = 500) for two cases: (1) from literature as in the GSA, and (2), from the distributions obtained after parameter estimation. Model simulations were performed in the Matlab environment (The Mathworks Inc., Natick, USA).

### 2.6. Parameter estimation procedure

The objective of the parameter estimation procedure was to sequentially estimate kinetic parameters that describe: (1) rates of hydrolysis, heterotrophic denitrification and heterotrophic oxygen consumption; (2) dynamics of nitrogen oxides (including NO and $N_2O$) under anoxia and endogenous conditions; (3) $NO_2^-$ oxidation; (4) $NH_4^+$ oxidation; (5) $N_2O$ production associated with $NH_4^+$ oxidation. The error function for problem minimization was defined as:

$$\mathrm{RMNSE} = \sum_{k}^{m}\sum_{j}^{n}\frac{\mathrm{RMSE_j}}{\bar{y}_{\mathrm{obs,j}}}; \qquad \mathrm{RMSE_j} = \sqrt{\frac{\sum_i^p\left(y_{\mathrm{sim,i}} - y_{\mathrm{obs,i}}\right)^2}{p}}$$

Where $m$ is the number of experiments in one scenario (e.g. 2 experiments where $NH_4^+$ is added: Scenario C), $n$ the number of data series in one experiment (e.g. NO, $N_2O$), $p$ the number of experimental points in each data series, $y_{sim,i}$ the model prediction and $y_{obs,i}$ the experimental data at time $i$. Parameters describing the elemental biomass composition (e.g. $i_{NXB}$), yield and temperature coefficients were fixed at default values and not subject to estimation. Newly estimated parameters were fixed at their best-fit estimate for the following parameter estimation step.

### 2.7. Validation of model response and parameter estimates

To test the validity of the model response (i.e., the adequacy of model to predict the observed data points) the interdependency of residuals ($y_{sim,i}$ - $y_{obs,i}$) was analysed by autocorrelation for different lag times (Cierkens et al., 2012). The quality of the model fit was evaluated via correlation coefficients ($R^2$) and more rigorously by an F-distribution test, with the hypothesis of a linear regression with simultaneous unit slope and zero intercept (Haefner, 2005). The identifiability of a parameter subset *K* was evaluated by a collinearity index ($\gamma_K$), which quantifies the near-linear dependence of local sensitivity functions. A collinearity index higher than 15 indicates a poorly identifiable parameter subset *K* (Brun et al., 2002). Approximate confidence regions were calculated following $J_{crit} = J_{opt}\left(1 + \frac{p}{N_{data}-p}F_{\alpha,p,N_{data}-p}\right)$ (Beale, 1960).

### 2.8. Case study: nitrification-denitrification cycle

To study the sensitivity of $N_2O$ and NO emissions during $NH_4^+$ removal with AS biomass a nitrification/denitrification cycle in a sequencing-batch reactor was simulated (**S-II**). Initially, $NH_3$ was added as a pulse (30 $mgN_{tot}$/L), consumed during 2 h of constant aeration, followed by 0.5 h anoxic period with excess carbon dosage (200 mgCOD/L). To investigate the effect of DO concentration on $N_2O$ emissions the aeration rates ($k_La_{O2}$) were adjusted to attain DO levels of approximately 0.5 and 2 mg/L respectively. The biomass content was set to 4 gTSS/L and stripping for $N_2O$ and NO was calculated from diffusivity corrections (Garcia-Ochoa and Gomez, 2009). A global sensitivity analysis was performed when removal rates were stable within one cycle (SRC method).

## 3. Results

### 3.1. Sensitivity analysis on the nitrification/denitrification case study

Simulation results for the nitrification/denitrification case study were used to investigate the main processes driving $N_2O$ production. The majority of $N_2O$ was produced and emitted during the aerobic part of the cycle at both low and high aeration rates, when $NH_4^+$ oxidation occurs. The highest ranked parameters were associated to DO and $NO_2^-$ sensitivity. The GSA ranking showed that at high DO the two most sensitive parameters correspond to AOB, followed by NOB and heterotrophic $NO_2^-$ reduction. At low DO all three microbial groups share high sensitivity (**S-II**). Overall, the results highlight the importance of interactions between AOB, NOB, and HB, on the $N_2O$ production from AS biomass during aerobic $NH_4^+$ oxidation. After identifying the key parameters activity tests can be designed accordingly.

### 3.2. Experimental results

**Biomass activity tests**

During Scenario A experiments heterotrophic denitrification and hydrolysis of biomass decay products were monitored. Under anoxic conditions the hydrolytic processes released $NH_4^+$ into the bulk (0.16 mg$NH_4^+$-N/gVSS/h). Simultaneously, biodegradable carbon was released and heterotrophic denitrification of $NO_3^-$, $NO_2^-$ and $N_2O$ was measured individually at excess electron acceptor concentrations (1.5, 2.5 and 4.7 mgN/gVSS/h respectively) (**Figure 1A**). The specific denitrification rates were significantly higher in the presence of excess electron donor (mix of C-sources) compared to endogenous conditions (7, 6.2 and 12 mgN/gVSS/h). The maximum $N_2O$ reduction rate varied 3-fold in the pH range 6.5 - 9, with a maximum at around pH = 8 (**Figure 1B**). Under endogenous aerobic conditions, a positive oxygen uptake rate was constantly measured, reflecting respiration of the biodegradable carbon and $NH_4^+$ released during hydrolysis (**S-V**). Addition of external biodegradable C-source increased the oxygen consumption rate compared to endogenous conditions (35 vs 4.5-7 mgCOD/gVSS/h).

Scenario C experiments were designed to study oxygen uptake associated with nitrification and started at excess DO (> 30mg/L). The oxygen uptake showed a dynamic response, increasing from a positive baseline up to a maximum rate after the N pulse , and decreasing until either the N-species ($NH_4^+$, $NO_2^-$) or DO reached limiting conditions ($OUR_{max,NH4}$ = 31 mg$O_2$/gVSS.h, $OUR_{max,NO2}$ = 18 mg$O_2$/gVSS/h) (**Figure 1C,D**).

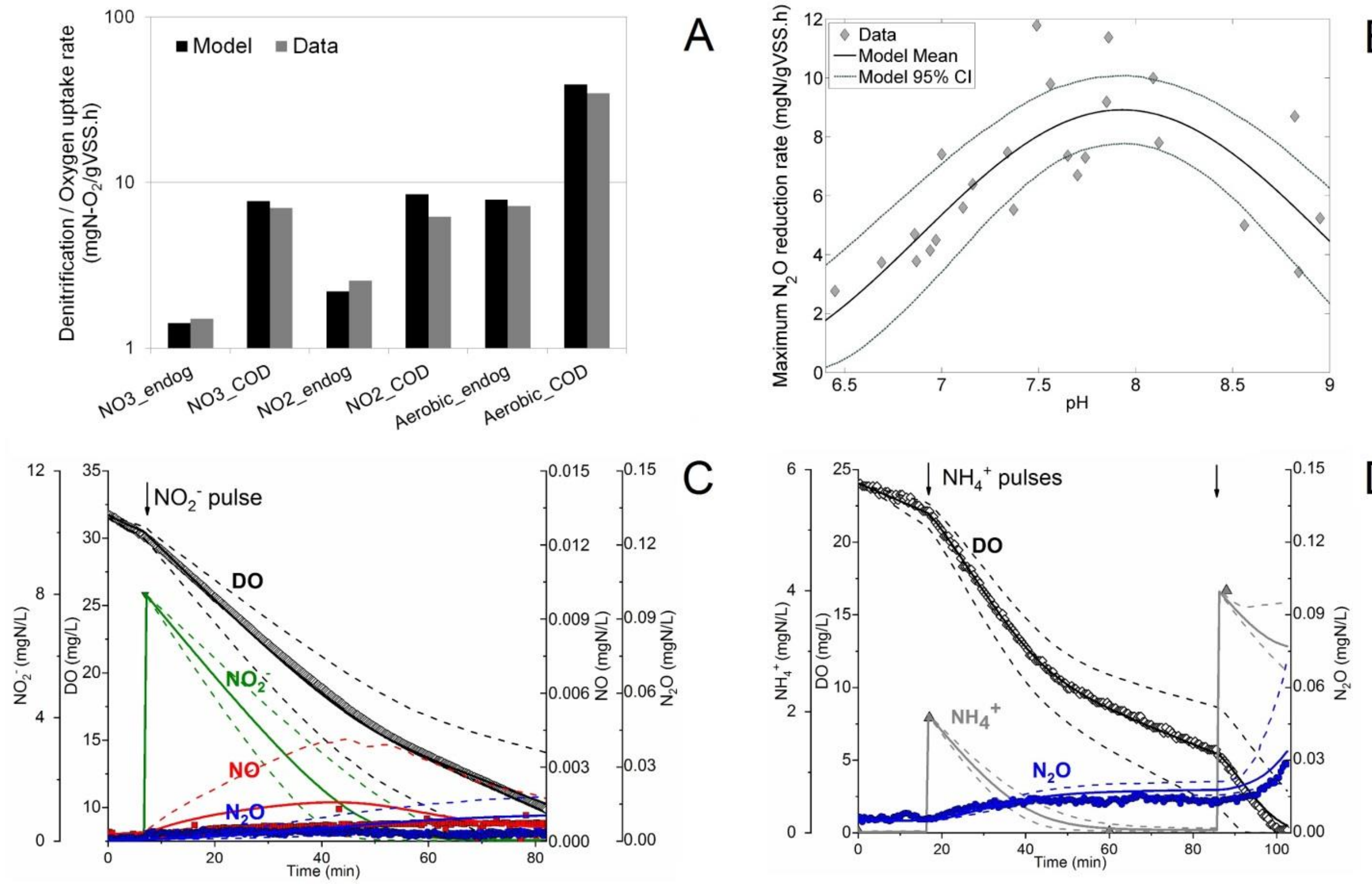

**Figure 1 –** Experimental and modelling results for heterotrophic processes (A, Scenario A), Anoxic $N_2O$ reduction dependency on pH (B, Scenario A), Aerobic $NO_2^-$ oxidation (C, Scenario C), Aerobic $NH_4^+$ oxidation (D, Scenario C). Experimental data (markers), best-fit model simulations (solid lines), 95% confidence intervals (dashed lines).

## $N_2O$ dynamics during respirometric assays

During Scenario B experiments under anoxic conditions, NO and $N_2O$ transiently accumulated after $NO_2^-$ and $NO_3^-$ pulses (**S-V**), but did not accumulate after sole $NH_4^+$ addition. After a $NO_3^-$ pulse NO concentration increased until reaching a maximum, followed by a steady decrease. The same trend was observed for $N_2O$ but delayed with respect to NO; $N_2O$ accumulation stopped after NO disappeared, followed by a continuous decrease. Similar NO and $N_2O$ dynamics was observed after $NO_2^-$ addition (**Figure 2A,B**), indicating NO as $N_2O$ precursor during heterotrophic denitrification.

In Scenario C experiments neither NO nor $N_2O$ accumulated in the bulk during $NO_2^-$ oxidation (**Figure 1C**). $NH_4^+$ oxidation lead to a low NO accumulation compared to the simultaneous and higher $N_2O$ increase (**Figure 1D, 2C**). The sole addition of $NH_2OH$, an intermediate of $NH_4^+$ oxidation to $NO_2^-$, yielded the highest NO and $N_2O$ accumulation rates ($N_2O$ accumulation rates ≈ 0 – 0.05 – 0.2 mgN/gVSS/h for $NO_2^-$, $NH_4^+$ and $NH_2OH$) (**Figure 2C**). Irrespective of the N-species being oxidized, at the onset of anoxia NO and $N_2O$ concentrations increased. First NO, and then $N_2O$, reached a maximum followed by a steady decrease, in a similar pattern to that observed after $NO_2^-$ or $NO_3^-$ pulses in anoxic experiments (Scenario B). The NO and $N_2O$ decrease indicate net consumption rates, but simultaneous production/consumption could still occur.

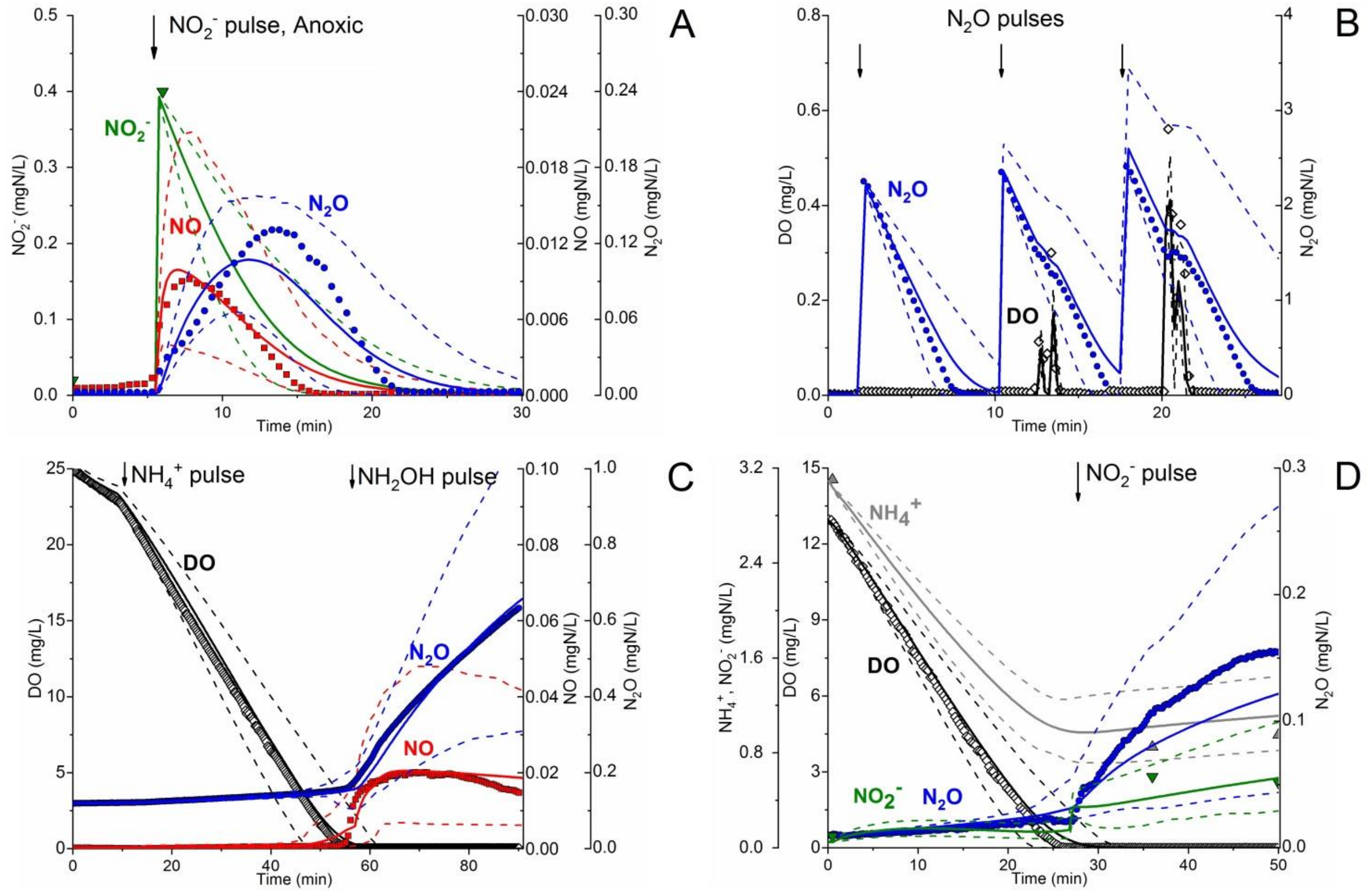

**Figure 2 –** Experimental and modelling results obtained during parameter estimation (A, Scenario A; B, Scenario B; C, Scenario C) and after parameter estimation (D, Scenario C). NO and $N_2O$ dynamics after anoxic $NO_2^-$ pulse under endogenous conditions (A). $N_2O$ consumption in the absence and presence of DO (DO pulses at t = 13, 21 min) (B). Oxygen consumption, NO and $N_2O$ dynamics after $NH_4^+$ pulse addition (t = 10 min) (C). Oxygen consumption and $N_2O$ dynamics after $NH_4^+$ pulse addition (D). Experimental data (markers), best-fit simulations (solid lines), 95% confidence intervals for the uncertainty of all model parameters (dashed lines).

### 3.3. Modelling results

The applicability of the NDHA model to describe the observed data was examined. First, the release of $NH_4^+$ and biodegradable organic carbon was studied under aerobic and anoxic conditions. The model simulated the anoxic $NH_4^+$ release from hydrolysis ($R^2$ = 0.97), and $\mu_{HB}$ was estimated to fit the oxygen uptake measured during Scenario A experiments (**Table 2**). Then, the maximum denitrification rates on $NO_3^-$ and $NO_2^-$ ($\mu_{NAR}$, $\mu_{NIR}$) were estimated from experiments in Scenario A). With the $N_2O$ dataset the pH-dependency of the $N_2O$ reduction was fitted using a sinusoidal model (Park et al., 2007) ($w_{NOS}$, $pH_{opt.NOS}$) while the substrate affinity for $N_2O$ reduction ($K_{HB.N2O}$) could be estimated from the $N_2O$-limited part of individual experiments from Scenario A. The NO and $N_2O$ concentrations from experiments in Scenario B, under endogenous respiration, were sensitive to changes in the affinity to biodegradable carbon for $NO_2^-$, NO and $N_2O$ reduction (**S-III**). Hence, the NO and $N_2O$ datasets were used to identify three of the affinities that were within the top sensitive parameters ($K_{HB.S.NIR}$, $K_{HB.S.NOR}$, $K_{HB.S.NOS}$, $\gamma < 15$).

The oxygen consumption rate is a measured variable in all experiments from Scenario C and in the NDHA model structure it is the sum of several processes: nitritation, nitratation, heterotrophic aerobic growth, decay and hydrolysis. The experimental design allowed the independent and sequential estimation of the interferences of these processes on AOB activity (**Table 2**). First, the substrate affinity and maximum growth rate of $NO_2^-$ oxidation ($K_{NOB.HNO2}$, $\mu_{NOB}$) were estimated ($\gamma < 7$). Then, $NH_4^+$ oxidation was described by estimating the substrate affinity and maximum growth rate of the first nitritation step ($K_{AOB.NH3}$, $\mu_{AOB.AMO}$). The NDHA model considers $NH_3$ as true substrate for AOB, and at lower pH levels $NH_3$ oxidation slows down, increasing the information content of the experiment compared to the same $NH_4^+$-N pulse at higher pH. Hence, the ammonia affinity for AOB ($K_{AOB.NH3}$ = 7 μgN/L) could be estimated from $NH_4^+$ pulses (2-3 mgN/L) at low pH as the DO sensitivity to changes in $K_{AOB.NH3}$ increases at lower pH levels (**S-III**).

After heterotrophic denitrification, nitrite oxidation, and ammonium oxidation had resulted in a good fit, parameters associated to AOB-driven $N_2O$ production were estimated in from NO and $N_2O$ data. The contribution of the NN pathway was estimated from the isolated $N_2O$ production during $NH_4^+$ oxidation at high DO, as $N_2O$ and NO were mostly sensitive to parameters associated to the NN pathway ($\varepsilon_{AOB}$, $\eta_{NOR}$) (**S-III**). The NO and $N_2O$ production observed at the onset of anoxia during $NH_4^+$ oxidation would correspond to the combined ND and HD contributions (**Figure 2C,D**). Hence, from experiments in Scenario C parameters describing the NN and ND pathways, $\varepsilon_{AOB}$, $\eta_{NIR}$ and $\eta_{NOR}$, could be identified ($\gamma_{NO,N2O} < 15$).

Based on the overall good fit of model predictions and experimental data the NDHA model described the dynamics of the measured DO and N-species ($R^2 \geq 0.94$, F-test = 1 for 10/11 datasets, **S-IV**). A total of 17 parameters were estimated sequentially with bounded approximate confidence regions indicating good identifiability (CV < 25%). Almost all the data points were within the 95% confidence interval of model predictions, considering the uncertainty of all model parameters, which indicates a good model description of the dataset (**S-IV**).

**Table 2** – Best-fit values for the parameters estimated (at 25 °C).

| Parameter | Units | Value | Scen. | Parameter | Unit | Value | Scen. | Parameter | Unit | Value | Scen. |
|---|---|---|---|---|---|---|---|---|---|---|---|
| $pH_{opt.nosZ}$ | ( - ) | 7.9 ± 0.1 | (A) | $K_{HB.S.NIR}$ | mgCOD/L | 4.3 ± 0.69 | (B) | $\mu_{AOB.AMO}$ | $d^{-1}$ | 0.86 ± 0.02 | (C) |
| $w_{nosZ}$ | ( - ) | 2.2 ± 0.2 | (A) | $K_{HB.S.NOR}$ | mgCOD/L | 5.3 ± 0.83 | (B) | $K_{AOB.NH3}$ | µgN/L | 7.00 ± 1.17 | (C) |
| $K_{HB.N2O}$ | mgN/L | 0.078 ± 0.020 | (A) | $K_{HB.S.NOS}$ | mgCOD/L | 4.1 ± 0.40 | (B) | $\varepsilon_{AOB}$ | ( - ) | 0.0031 ± 0.00 | (C) |
| $\mu_{HB.NAR}$ | $d^{-1}$ | 1.71 ± 0.11 | (A) | $\mu_{HB}$ | $d^{-1}$ | 7.23 ± 0.16 | (A) | $\eta_{NIR}$ | ( - ) | 0.22 ± 0.01 | (C) |
| $\mu_{HB.NIR}$ | $d^{-1}$ | 1.11 ± 0.07 | (A) | $\mu_{NOB}$ | $d^{-1}$ | 1.51 ± 0.07 | (C) | $\eta_{NOR}$ | ( - ) | 0.36 ± 0.02 | (C) |
| $\mu_{HB.NOS}$ | $d^{-1}$ | 1.17 ± 0.02 | (A) | $K_{NOB.HNO2}$ | µgN/L | 0.027 ± 0.006 | (C) | | | | |

Single $NH_2OH$ pulses were not considered for parameter estimation as the electron flow during $NH_2OH$ oxidation in the AOB metabolism differs from that during $NH_4^+$ oxidation (two out of the four electrons are shuttled back from HAO to AMO). Hence, $NH_2OH$ pulses are not representative of $NH_4^+$-oxidation driven $N_2O$ production. Also, the contribution of $NH_2OH$ to abiotic NO/$N_2O$ production was not significant after $NH_2OH$ pulses for experiments in Scenario C (**S-V**).

The analysis of residuals from best-fit simulations after parameter estimation with the complete datasets showed a high autocorrelation of the residuals for some experiments. The DO and $N_2O$ datasets were down-sampled by lowering the data acquisition frequency down to non-autocorrelated values (Cierkens et al., 2012). In Scenario C, the DO dataset used to estimate the parameter subset $K_{AOB.NH3}$ and $\mu_{AOB.AMO}$ was down-sampled fourteen-fold to correct autocorrelation (from 1230 to 96 data points). With the updated dataset the best-fit estimates did not change significantly (0.7 and 4.2% variation between datasets), but the uncertainty increased almost four-fold (**S-IV**).

### 3.4. Model evaluation

The predictive ability of the calibrated NDHA model was evaluated on a set of batch experiments where the AS biomass from the same WWTP was subject to varying N pulses under constant aeration (May-June 2012) (Domingo-Félez et al., 2017b). The model was evaluated in Scenarios D and E against DO, $NH_4^+$, $NO_2^-$ and $N_2O$ data. Only the oxygen mass transfer coefficient ($k_La_{O2}$) was tuned as the aeration in the respirometric assays from Scenarios A, B, C was null. Overall, the model captured the trends of DO, main N-species and liquid $N_2O$ without any need for parameter value modification ($R^2_{avg}$ for DO = 0.98; $NH_4^+$ = 0.99; $NO_2^-$ = 0.84; $N_2O$ = 0.80). Only the $N_2O$ residuals ($y_{sim,i}$ - $y_{obs,i}$) did not pass the F-distribution test ($F_{N2O}$ = 0) (**Figure 3**).

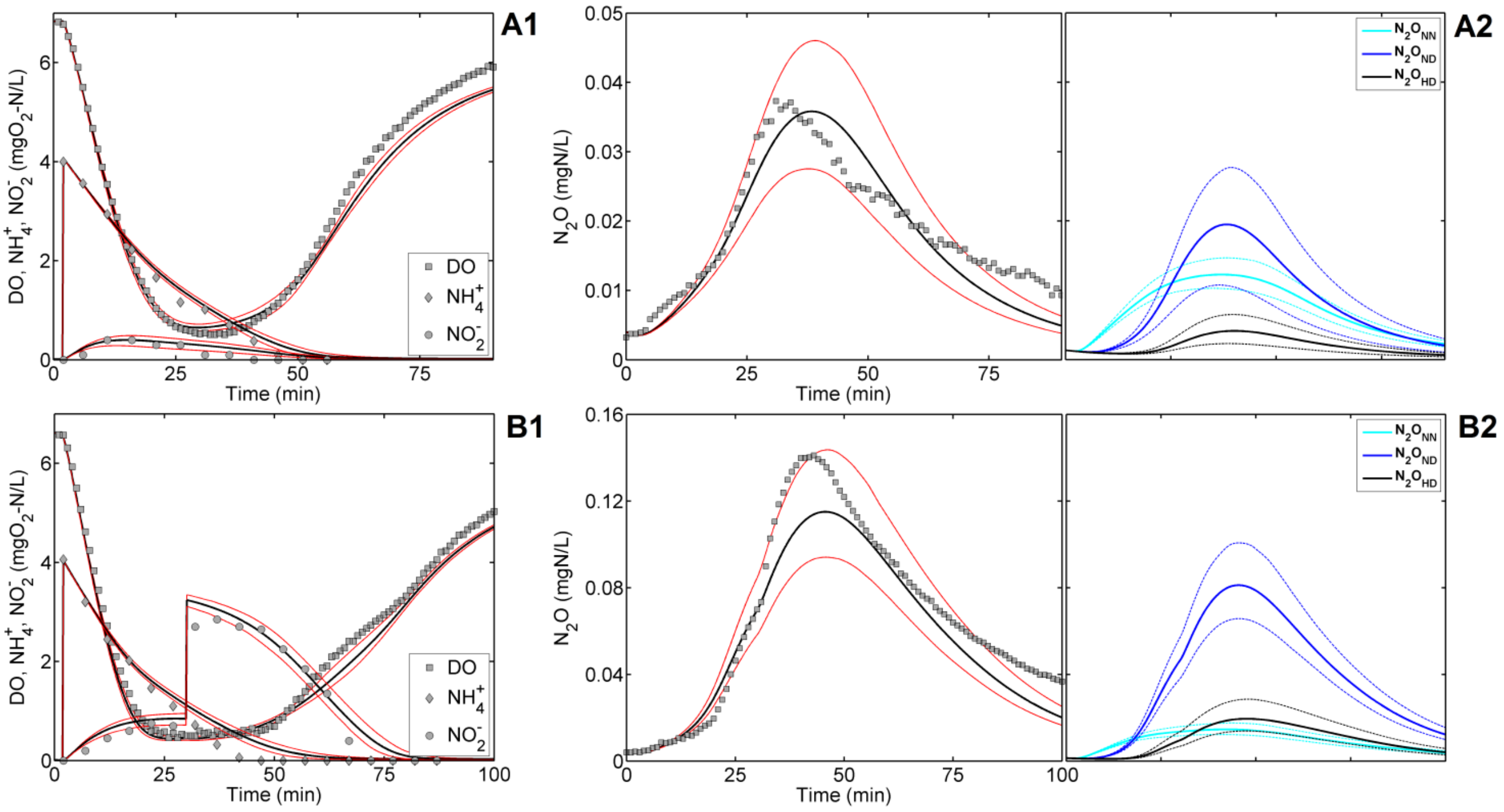

**Figure 3 –**Model validation results from Scenario D. Effect of $NO_3^-$ pulse (A1,A2) and $NO_2^-$ pulse (B1,B2) during aerobic $NH_4^+$ oxidation. Main substrates: DO, $NH_4^+$, $NO_2^-$ (A1, B1), $N_2O$ and $N_2O$ pathway contributions (A2, B2). Experimental results (markers), best-fit simulations (black lines), 95% confidence intervals (red lines). Nitrifier nitrification (cyan), nitrifier denitrification (blue), and heterotrophic denitrification (black) pathway contributions; mean (solid lines) and 95% confidence intervals (dashed lines). Confidence intervals of uncertainty from the estimated parameters only (Table 2).

Higher $NH_4^+$ pulses yielded a higher $N_2O$ fraction (Scenario D1-D4: 0.6 - 1.7 - 2.5 - 3.2% $N_2O/NH_4^+{}_{rem}$) as more $NH_4^+$ oxidation occurred at low DO, thus promoting the contribution of denitrification pathways (**Figure 4**). Addition of a $NO_2^-$ pulse (Scenario E-3) increased the fraction of $N_2O$ produced compared to a $NO_3^-$ pulse or no pulse (Scenario E-1, E-2).

| | Pulse (mgN/L) | | | $N_2O$ emission factor | $N_2O$ pathway contribution | | |
|---|---|---|---|---|---|---|---|
| Scen. | $NH_4^+$ | $NO_2^-$ | $NO_3^-$ | $N_2O$ / $NH_{4,rem}$ | NN (%) | ND (%) | HD (%) |
| D-1 | 2.3 | | | 0.6% | 57% | 33% | 10% |
| D-2 | 3.6 | | | 1.7% | 22% | 57% | 21% |
| D-3 | 4.7 | | | 2.5% | 19% | 60% | 21% |
| D-4 | 5.5 | | | 3.2% | 13% | 64% | 23% |
| E-1 | 3.9 | | | 0.8% | 43% | 44% | 13% |
| E-2 | 4.0 | | 2.0 | 0.8% | 41% | 47% | 12% |
| E-3 | 4.1 | 2.5 | | 1.7% | 17% | 66% | 17% |

**Figure 4 –** Model evaluation results for continuously aerated batch tests for different $NH_4^+$, $NO_2^-$, $NO_3^-$ pulse additions: $N_2O$ emission factor and $N_2O$ pathway contributions.

In terms of pathway contributions to the total $N_2O$ pool, model predicts that after addition of an $NH_4^+$ pulse at high DO the NN pathway has the largest contribution (**Figure 3**). When $NO_2^-$ accumulates in the bulk and DO reaches low concentrations (DO < 0.5 mg/L) the ND and HD contributions increase. Even though the HD pathway had a positive contribution to the $N_2O$ pool (no net $N_2O$ consumption) it was significantly lower than the ND pathway (**Figure 4**).

$N_2O$ emissions from AS biomass during aerobic $NH_4^+$ oxidation simulations with best-fit estimate parameters were predicted for a wider range of DO (0.2 – 4 mg/L) and $NO_2^-$ (0 – 1.4 mgN/L) at excess $NH_4^+$. The $N_2O$ emission factor and individual pathway contributions to the total $N_2O$ pool at pseudo-steady state are shown in **Figure 5**. The NN pathway contributes most at the lowest $NO_2^-$ and highest DO (98%), and the least at high $NO_2^-$ and low DO (3%). The ND and HD pathways showed similar trends and opposite compared to the NN pathway, with maximum contributions of 72% and 43% respectively. The $N_2O$ emission factor ranged between 0.45 and 12.3% with a very sharp increase towards low DO and high $NO_2^-$. For example, at DO = 2 mg/L, with increasing $NO_2^-$ concentrations the emission factor increased from 0.45 to 1.26%.

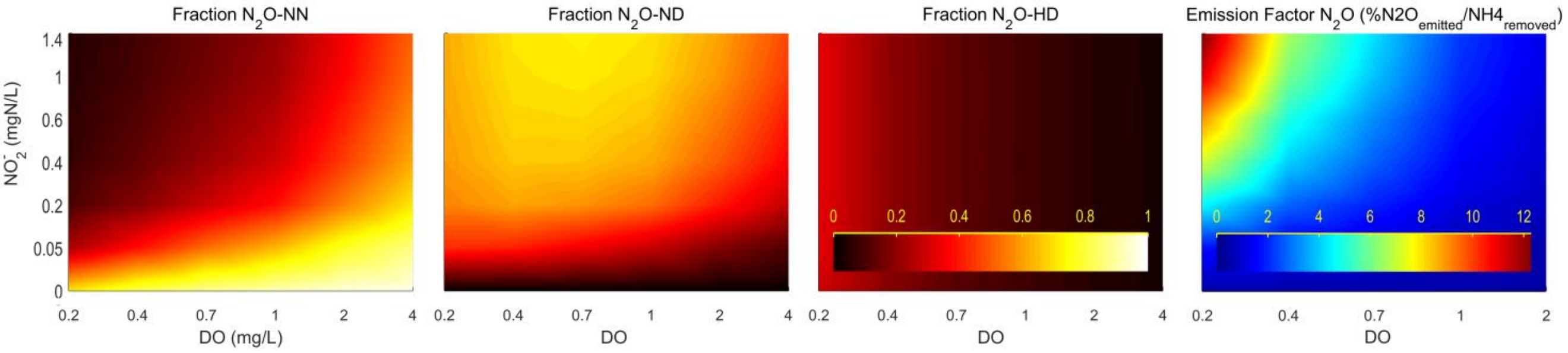

**Figure 5 –** Model evaluation at varying $NO_2^-$ and DO concentrations during excess $NH_4^+$ removal (pH = 7.2). From left to right: Pathway contributions to total $N_2O$ pool NN, ND, HD; $N_2O$ emission factor.

## 4. Discussion

### 4.1. N-dynamics in respirometric assays: experimental and modelling results

Under endogenous conditions the lowest denitrification rate corresponded to $NO_3^-$ reduction, followed by $NO_2^-$ and $N_2O$ reductions (**Figure 1**). A limiting electron flow could explain these differences as more reduced $NO_x^-$ compounds require fewer electrons to produce $N_2$. To avoid a biased denitrification rate due to a single carbon source (Lu et al., 2014), a mixture of three carbon sources was considered representative of a complex wastewater (20% formate, 40% acetate, 40% propionate COD-based). Overall, the denitrification rates observed were in the low range of literature values (**S-V**). The differences in the carbon usage by each reduction step can be linked to the high modularity of the heterotrophic denitrifying community (Graf et al., 2014). After model fitting, the reduction factors for the denitrification steps compared to the aerobic growth rate are similar to reported values (Hiatt and Grady, 2008).

$N_2O$ is typically found at very low concentrations due to stripping and/or simultaneous production/consumption processes. Hence, the only biological $N_2O$ consumption process could be isolated by only providing $N_2O$ at high concentrations as the only N-source (> 1mgN/L). pH changes significantly affected the maximum $N_2O$ consumption rate, in a similar fashion to that found by Pan *et al.*, (2012) for a methanol-fed denitrifying culture. By including a sinusoidal function the model captured this trend (**Figure 1**). The substrate affinity was independently estimated as $\mu_{HB.NOS}$ had been previously fixed ($K_{HB.N2O}$ = 0.078mgN/L) and, differently from Pan *et al.*, (2012), the individual $K_{HB.N2O}$ values estimated from each experiment did not correlate with pH ($R^2$ = 0.062, n = 12, **S-IV**). Liquid $N_2O$ concentrations are commonly measured at similar or lower values than $K_{HB.N2O}$, indicating that $N_2O$ consumption is a substrate limited process and specific experiments must be designed to estimate it (Kampschreur et al., 2008; Lindblom et al., 2015).

Moreover, the reported experimental design benefited from NO measurements compared to only $N_2O$ measurements as it allowed: (1) identification of specific parameters (e.g. $K_{HB.S.NIR}$) and (2) lowering the estimated uncertainty of $\varepsilon_{AOB}$ and $\eta_{NOR}$ 4 and 9-fold respectively (**S-IV-1**). The three reduction factors associated to AOB-driven $N_2O$ production could be estimated: $\varepsilon_{AOB}$ (NN), $\eta_{NIR}$ (ND) and $\eta_{NOR}$ (NN + ND). Direct comparison of the estimated values to other studies is not possible due to differences in model structure. However, $\varepsilon_{AOB}$ (0.0031) and $\eta_{NIR}$ (0.22) were in the same range of other reduction factors from 2-pathway AOB models ($\eta_{NN}$ = 0.0007 - 0.12, $\eta_{NIR}$ = 0.03 – 0.48) (**S-I**).

### 4.2. Applicability of the NDHA model to mixed liquor biomass

After model validation the simulated nitrification/denitrification cycle showed that almost all the $N_2O$ was emitted during the aerated phase, as during the anoxic phase $N_2O$ was consumed with excess COD (data not shown). The simulated cycle at low DO yielded a higher $N_2O$ emission factor as compared to that at higher DO (4.6 and 1.2% respectively), in agreement with other SND systems (Hu et al., 2010; Tallec et al., 2006). The $N_2O$ emission factors are comparable with those reported by Wunderlin *et al.* (2012) after $NH_4^+$ pulses at the same DO levels (3.8 and 2%). Also, in a full-scale activated sludge plant the emission factor varied between 0.69 and 3.5% for a range of DO and $NO_2^-$ conditions (Ni et al., 2015).

The contribution of the NN pathway increased with DO, while the two denitrification pathways, ND and HD, decreased (**Figure 6**). For both DO levels the ND contribution was significantly larger than the HD contribution (64 vs 17%, 42 vs 7% respectively). Similarly, in other modelling studies an anaerobic/oxic/anoxic process showed that at DO < 1 mg/L the ND pathway was also the main contributor to the total $N_2O$ pool, followed by NN and HD (Ding et al., 2016). However, separating the ND and HD contributions from $NH_4^+$ oxidation processes with AS biomass is not trivial (Domingo-Félez et al., 2017b). Even at minimum C/N ratios the high heterotrophic abundance compensates for the low electron donor concentrations, yielding similar ND and HD rates, both increasing with $NO_2^-$ and low DO concentrations. For example, others studies predicted a higher HD contribution compared to ND based on experimental observations (Hu et al., 2011; Wunderlin et al., 2012).

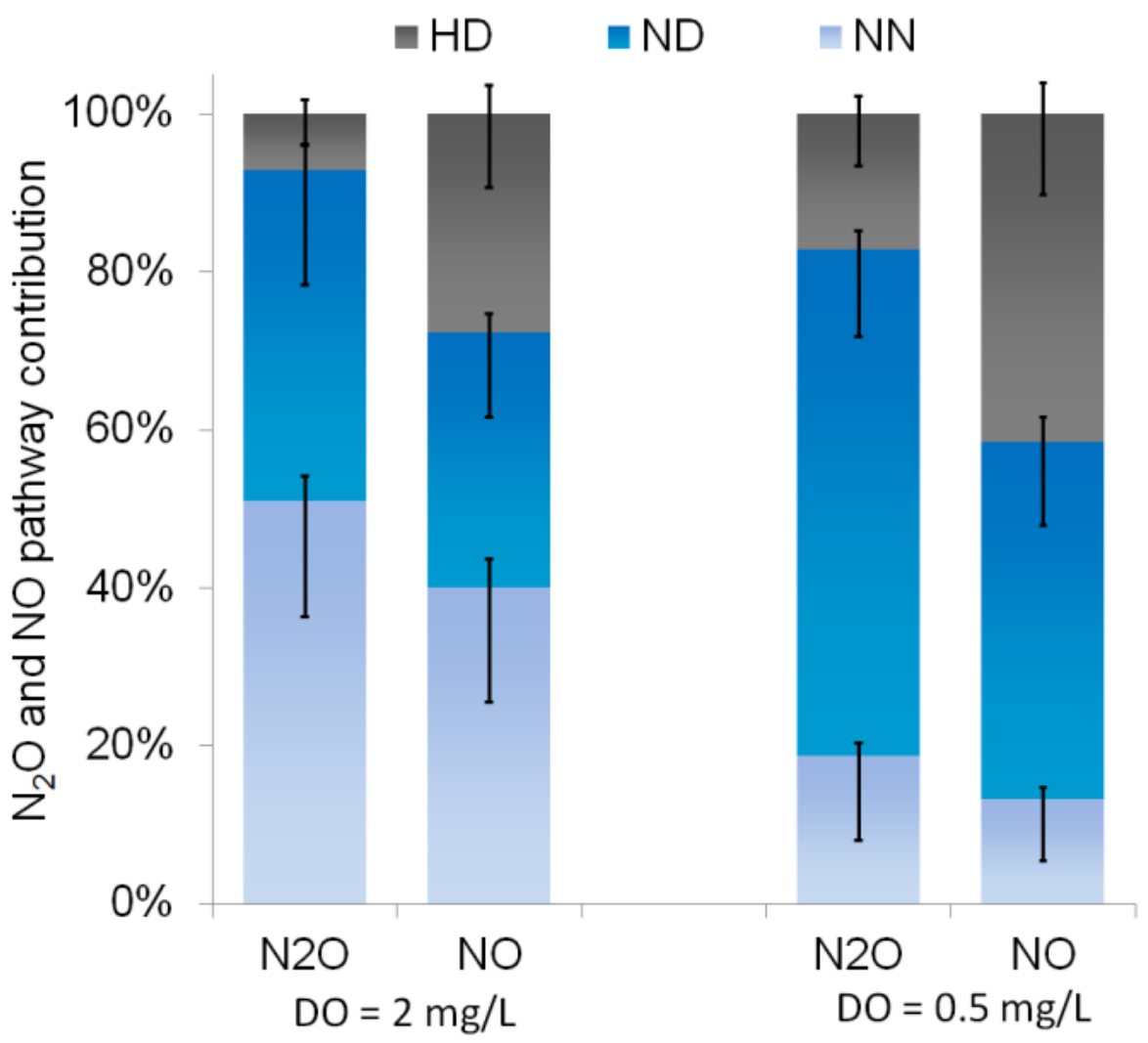

**Figure 6 –** $N_2O$ and NO pathway contribution for the nitrification/denitrification case study after model calibration. The error bars correspond to the standard deviation of the uncertainty from estimated parameters: (top bar) in this study, (bottom bar) default following (Sin et al., 2009), n = 500.

### 4.3. Role of nitric oxide on $N_2O$ models

$N_2O$ model predictions are evaluated in two steps: first, DO and main nitrogenous substrates (e.g. $NH_4^+$, $NO_2^-$, $NO_3^-$) and secondly $N_2O$. Models with similar $N_2O$ pathway structures cannot be discriminated only with $N_2O$ datasets, and additional measurements such as NO can improve model discrimination (Lang et al., 2017). However, only a few $N_2O$ models have been tested against NO (Spérandio et al., 2016).

In this study net NO and $N_2O$ production from $NH_4^+$ oxidation under aerobic conditions was significantly lower than at the onset of anoxia (**Figure 2**), as was also reported for pure AOB cultures and nitrifying systems (Kampschreur et al., 2008; Kozlowski et al., 2016; Wunderlin et al., 2012). The higher anoxic rates can be explained by the transient accumulation of $NH_2OH$, which has been suggested to act as electron donor for $NO_2^-$ reduction to $N_2O$ in a two-step process over NO (de Bruijn et al., 1995; Kester, 1997; Poth and Focht, 1985). The transient accumulation of NO at low DO conditions highlights the importance of $N_2O$ models including NO as an intermediate of the ND pathway. Simultaneously, incomplete heterotrophic denitrification also contributes to the observed net NO and $N_2O$ production.

The ratio between the substrate affinity of NO reductases, $K_{AOB.NO}/K_{HB.NO}$, is an important parameter of $N_2O$ models as it can shift the predicted contributions of the ND and HD pathways for the same overall $N_2O$ fit (Domingo-Félez et al., 2017b). However, direct estimation of NO affinity is difficult due to cellular toxicity (Schulthess and Gujer, 1996). In $N_2O$ models $K_{NO}$ values are typically assumed (Hiatt and Grady, 2008; Pan et al., 2013) and, when estimated, no identifiability metrics are reported (Schreiber et al., 2009). This could partially explain the high variability of reported NO affinities and $K_{AOB.NO}/K_{HB.NO}$ ratios considered in literature ($K_{HB.NO}$ = 0.00015-0.05 mgN/L and $K_{AOB.NO}$, 0.004-0.1mgN/L, $K_{AOB.NO}/K_{HB.NO}$ = 1 - 56) (Hiatt and Grady, 2008; Ni et al., 2011; Schreiber et al., 2009; Spérandio et al., 2016; Q. Wang et al., 2016). For example, in the study by Wang *et al.* (2016) the HD pathway has an NO affinity over 50 times higher than the NN pathway. The HD pathway could theoretically uptake NO produced by the NN pathway at a much higher rate and could underestimate the NN contribution to the total $N_2O$ pool. Based on current lack of knowledge and to avoid a preferential NO-consumption/$N_2O$-production pathway the NO affinity ratio of AOB and HB was set equal. Future $N_2O$ model calibrations would benefit from stable-isotope measurements (15N-18O labelling) to distinguish the pathway contribution by AOB and HB.

### 4.4. Role of hydroxylamine on $N_2O$ models

$NH_2OH$ is an important intermediate in the 2-step $NH_4^+$ oxidation to $NO_2^-$ as the electron source for the metabolism of AOB. Low $NH_2OH$ bulk concentrations have been reported in the liquid phase of pure AOB cultures and nitrifying systems ($NH_2OH$ < 0.1mgN/L) (Soler-Jofra et al., 2016; Yu and Chandran, 2010), indicating a quick turnover of $NH_2OH$. $N_2O$ models have adopted the 2-step $NH_4^+$ oxidation where $NH_2OH$ is the direct substrate for NO and $N_2O$ production (Ding et al., 2016; Ni et al., 2014, 2013, 2011; Pocquet et al., 2016). However, $NH_2OH$ predictions are not addressed. Parameter sets of some $N_2O$ models would overestimate $NH_2OH$ equilibrium concentrations during $NH_4^+$ oxidation as $\mu_{AMO} \geq \mu_{HAO}$ and low $NH_2OH$ affinities are considered ($K_{AOB.NH2OH}$ = 0.2 mgN/L this study, 0.7 – 2.4mgN/L literature) (Ding et al., 2016;

Ni et al., 2014, 2013, 2011; Pocquet et al., 2016). For more accurate $NH_2OH$ predictions a faster HAO process compared to AMO is necessary as it prevents high $NH_2OH$ accumulations. This is in agreement with the calibrated NDHA model where $\mu_{AMO} < \mu_{HAO}$ and $K_{AOB.NH2OH} < K_{AOB.NH4}$. The NDHA model predicted average $NH_2OH$ concentrations of 0.15 ± 0.09 mgN/L in Scenario C experiments and 0.13 ± 0.04 mgN/L in Scenario D experiments, in close agreement with measured $NH_2OH$ in a shortcut nitrification reactor of 0.06 mgN/L (Hu et al., 2018).

The higher $N_2O$ yield observed after $NH_2OH$ pulses compared to other N-species under aerobic conditions is in agreement with other studies (≈ 0.05 and 0.2 $mgN_2O$-N/gVSS/h for $NH_4^+$ and $NH_2OH$) (Kim et al., 2010; Terada et al., 2017; Wunderlin et al., 2012). The NDHA model structure can describe $N_2O$ production associated to aerobic $NH_4^+$ oxidation and aerobic/anoxic $NH_2OH$ oxidation. However, the high $N_2O$ production rate associated to aerobic $NH_2OH$ pulses (1-2 mgN/L) could not be predicted by the calibrated model. However, the $NH_2OH$ concentrations explored were very high and not representative of wastewater treatment operations, the aim of this study (Caranto et al., 2016; Terada et al., 2017). These results suggest that other $N_2O$ pathways, such as the associated to Cyt P460, could be responsible for the high $N_2O$ production observed during aerobic $NH_2OH$ oxidation (Kozlowski et al., 2016). Hence, the NDHA model does not describe individually all the co-occurring $N_2O$ pathways in the AOB metabolism. However, the relevant biological $N_2O$ production associated to wastewater treatment conditions is captured in the NN and ND processes.

### 4.5. $N_2O$ emissions: Parameter uncertainty propagation

During the past years sequential $N_2O$ dynamic modelling efforts have focused on improving model structures to capture the variety of $N_2O$ production phenomena and rates observed during wastewater treatment operations. From single to double AOB pathways, and in combination with HB, $N_2O$ models have increased in complexity (i.e. more processes and parameters) and improved in their best-fit capabilities. The confidence of model predictions is usually addressed [FOR WHAT???] in WWTP models but has not been studied for $N_2O$ models. The uncertainty in prediction of N-removing processes propagates into prediction of $N_2O$ production, increasing the uncertainty of $N_2O$ emissions compared to main N-species (**Figure 4**). Here, to evaluate the accuracy of the biological parameter estimates from respirometric assays the uncertainty of $N_2O$ emissions in the nitrification/denitrification case study was analysed following Sin *et al.*, (2009). Two cases were analysed: the calibrated parameter subset with the estimated uncertainty from Table 2, and using uncertainty classes as a reference (**S-II**). By comparing the two cases the uncertainty calculated with this calibration procedure was only 28% of that simulated with the reference. The calibrated NDHA model predicted for low and high DO an $N_2O$ emission factor of 4.6 ± 0.6% and 1.2 ± 0.1%, which corresponds to low coefficients of variation (9 and 12%) (**S-VI**). With the same uncertainty evaluation methodology, the calibrated NDHA model for a an AOB-enriched biomass showed an 8% variation of the N2O emission factor when only the estimated parameters were considered, but 40% when all the model parameters were considered uncertain (Domingo-Félez et al., 2017a). These results highlight the importance of evaluating the uncertainty of biological parameter estimates in $N_2O$ emissions, but unfortunately no other studies are available. We believe that future comparison of best-fit simulations together with their uncertainty (e.g. 95% CI) will improve calibration procedures for $N_2O$ models.

## 5. Conclusions

- In respirometric and batch assays $N_2O$ and NO production from mixed liquor biomass increased during $NH_4^+$ oxidation under low DO concentrations and the presence of $NO_2^-$.
- A model considering three biological $N_2O$ production pathways was calibrated and predicted the NO and $N_2O$ dynamics at varying $NH_4^+$, $NO_2^-$ and DO levels.
- In the $NH_4^+$ oxidation experiments used for model evaluation the NN pathway showed the largest contribution at high DO levels, while the ND and HD pathways increased and dominated the total $N_2O$ production at low DO and high $NO_2^-$ concentrations.
- The NDHA model could not predict $N_2O$ production during aerobic $NH_2OH$ oxidation. However, it is of minor relevance for wastewater treatment operations where $NH_2OH$ accumulates at very low levels.
- The uncertainty of $N_2O$ model predictions was evaluated and could be used in future studies to discriminate between calibration procedures.

**Software availability**

The MATLAB/SIMULINK code containing the implementation of the model is free upon request to the corresponding author.

**Acknowledgements**

This research was funded by the Danish Agency for Science, Technology and Innovation through the Research Project LaGas (12-132633). The authors have no conflict of interest to declare. Dr. Ulf Jeppsson (Lund University) is acknowledged for having provided the codes of the Benchmark Simulation Model no 2 from which this work was developed.

# Supplemental Information

**Title**

Modelling $N_2O$ dynamics of activated sludge biomass under nitrifying and denitrifying conditions: pathway contributions and uncertainty analysis.

**Author list**

Carlos Domingo-Félez, Barth F. Smets*

Department of Environmental Engineering, Technical University of Denmark, Miljøvej 115, 2800 Kgs. Lyngby, Denmark

* Corresponding author:

Barth F. Smets, Phone: +45 4525 1600, Fax: +45 4593 2850, E-mail: bfsm@env.dtu.dk

---

## S-I NDHA model

**Table S-I -1 –** Stoichiometric matrix and process rates of relevant processes of the NDHA model. Adapted from Domingo-Félez and Smets (2016).

| Component ▸ / Process ▾ | | 1 $S_S$ | 2 $S_{O2}$ | 3 $S_{NH3}$ | 4 $S_{NH2OH}$ | 5 $S_{HNO2}$ | 6 $S_{NO3}$ | 7 $S_{NO}$ | 8 $S_{N2O}$ | 9 $S_{N2}$ | 11 $X_{B,AOB}$ | 12 $X_{B,NOB}$ | 13 $X_{B,H}$ | 14 $X_S$ | 15 $X_I$ |
|---|---|---|---|---|---|---|---|---|---|---|---|---|---|---|---|
| ***AOB growth*** | | | | | | | | | | | | | | | |
| Aerobic_AMO | 1 | | -1.14 | -1 | 1 | | | | | | | | | | |
| Aerobic_HAO* | 2 | | | $-i_{NXB}$ | $-\frac{1}{Y_{AOB}}$ | | | $\frac{1}{Y_{AOB}}$ | | | 1 | | | | |
| Aerobic_HAO | 3 | | $-\left(\frac{2.29-Y_{AOB}}{Y_{AOB}}\right)$ | $-i_{NXB}$ | $-\frac{1}{Y_{AOB}}$ | $\frac{1}{Y_{AOB}}$ | | | | | 1 | | | | |
| Anox_A_NIR | 4 | | | | -1 | -3 | | 4 | | | | | | | |
| Anox_A_NOR | 5 | | | | -1 | | | -2 | 3 | | | | | | |
| ***NOB growth*** | | | | | | | | | | | | | | | |
| Aer_NOB_growth | 6 | | $-\left(\frac{1.14-Y_{NOB}}{Y_{NOB}}\right)$ | $-i_{NXB}$ | | $-\frac{1}{Y_{NOB}}$ | $\frac{1}{Y_{NOB}}$ | | | | | 1 | | | |
| ***HB growth*** | | | | | | | | | | | | | | | |
| Aerobic_H_growth | 7 | $-\frac{1}{Y_{HB}}$ | $-\left(\frac{1-Y_{HB}}{Y_{HB}}\right)$ | $-i_{NXB}$ | | | | | | | | | 1 | | |
| Anox_H_NAR | 8 | $-\frac{1}{Y_{HB}}$ | | $-i_{NXB}$ | | $\left(\frac{1-Y_{HB}}{1.14\cdot Y_{HB}}\right)$ | $-\left(\frac{1-Y_{HB}}{1.14\cdot Y_{HB}}\right)$ | | | | | | 1 | | |
| Anox_H_NIR | 9 | $-\frac{1}{Y_{HB}}$ | | $-i_{NXB}$ | | $-\left(\frac{1-Y_{HB}}{0.57\cdot Y_{HB}}\right)$ | | $\left(\frac{1-Y_{HB}}{0.57\cdot Y_{HB}}\right)$ | | | | | 1 | | |
| Anox_H_NOR | 10 | $-\frac{1}{Y_{HB}}$ | | $-i_{NXB}$ | | | | $-\left(\frac{1-Y_{HB}}{0.57\cdot Y_{HB}}\right)$ | $\left(\frac{1-Y_{HB}}{0.57\cdot Y_{HB}}\right)$ | | | | 1 | | |
| Anox_H_NOS | 11 | $-\frac{1}{Y_{HB}}$ | | $-i_{NXB}$ | | | | | $-\left(\frac{1-Y_{HB}}{0.57\cdot Y_{HB}}\right)$ | $\left(\frac{1-Y_{HB}}{0.57\cdot Y_{HB}}\right)$ | | | 1 | | |
| ***Lysis*** | | | | | | | | | | | | | | | |
| AOB | 12 | | | $i_{NXB}-f_I.i_{NXI}-(1-f_I).i_{NXS}$ | | | | | | | -1 | | | $1-f_{XI}$ | $f_{XI}$ |
| NOB | 13 | | | $i_{NXB}-f_I.i_{NXI}-(1-f_I).i_{NXS}$ | | | | | | | | -1 | | $1-f_{XI}$ | $f_{XI}$ |
| HB | 14 | | | $i_{NXB}-f_I.i_{NXI}-(1-f_I).i_{NXS}$ | | | | | | | | | -1 | $1-f_{XI}$ | $f_{XI}$ |
| ***Hydrolysis*** | | | | | | | | | | | | | | | |
| Aerobic | 15 | 1 | | $i_{NXS}$ | | | | | | | | | | -1 | |
| Anoxic | 16 | 1 | | $i_{NXS}$ | | | | | | | | | | -1 | |
| Anaerobic | 17 | 1 | | $i_{NXS}$ | | | | | | | | | | -1 | |

| *Process* ▼ | | Process Rate (g·m[-3]·min[-1]) |
|---|---|---|
| ***AOB growth*** | | |
| Aerobic_AMO | *1* | $\mu_{AMO}^{AOB}\cdot\frac{S_{O2}}{S_{O2}+K_{O2_AMO}^{AOB}}\cdot\frac{S_{NH3}}{S_{NH3}+K_{NH3}^{AOB}+S_{NH3}^{2}/K_{I_NH3}^{AOB}}\cdot\frac{K_{I_HNO2}^{AOB}}{S_{HNO2}+K_{I_HNO2}^{AOB}}\cdot X_{AOB}$ |
| Aerobic_HAO* | *2* | $\mu_{HAO}^{AOB}\cdot\varepsilon_{AOB}\cdot\frac{S_{NH2OH}}{S_{NH2OH}+K_{NH2OH}^{AOB}}\cdot X_{AOB}$ |
| Aerobic_HAO | *3* | $\mu_{HAO}^{AOB}\cdot(1-\varepsilon_{AOB})\cdot\frac{S_{O2}}{S_{O2}+K_{O2_HAO}^{AOB}}\cdot\frac{S_{NH2OH}}{S_{NH2OH}+K_{NH2OH}^{AOB}}\cdot X_{AOB}$ |
| Anox_A_NIR | *4* | $\mu_{HAO}^{AOB}\cdot\eta_{NIR}\cdot\frac{K_{I_O2}^{AOB}}{S_{O2}+K_{I_O2}^{AOB}}\cdot\frac{S_{NH2OH}}{S_{NH2OH}+K_{NH2OH_ND}^{AOB}}\cdot\frac{S_{HNO2}}{S_{HNO2}+K_{HNO2}^{AOB}}\cdot X_{AOB}$ |
| Anox_A_NOR | *5* | $\mu_{HAO}^{AOB}\cdot\eta_{NOR}\cdot\frac{S_{NH2OH}}{S_{NH2OH}+K_{NH2OH_ND}^{AOB}}\cdot\frac{S_{NO}}{S_{NO}+K_{NO_ND}^{AOB}}\cdot X_{AOB}$ |
| ***NOB growth*** | | |
| Aer_NOB_growth | 6 | $\mu_{NOB}\cdot\frac{S_{O2}}{S_{O2}+K_{O2}^{NOB}}\cdot\frac{S_{HNO2}}{S_{HNO2}+K_{HNO2}^{NOB}+S_{HNO2}^{2}/K_{I_HNO2}^{NOB}}\cdot\frac{K_{I_NH3}^{NOB}}{S_{NH3}+K_{I_NH3}^{NOB}}\cdot X_{NOB}$ |
| ***HB growth*** | | |
| Aerobic_HB_growth | 7 | $\mu_{HB}\cdot\frac{S_{O2}}{S_{O2}+K_{O2}^{HB}}\cdot\frac{S_{NH4}}{S_{NH4}+K_{NH4}^{HB}}\cdot\frac{S_{S}}{S_{S}+K_{S}^{HB}}\cdot X_{HB}$ |
| Anox_HB_NAR | 8 | $\mu_{NAR}^{HB}\cdot\eta_{HD}\cdot\frac{K_{I_O2_NAR}^{HB}}{S_{O2}+K_{I_O2_NAR}^{HB}}\cdot\frac{S_{S}}{S_{S}+K_{S_NAR}^{HB}}\cdot\frac{S_{NH4}}{S_{NH4}+K_{NH4}^{HB}}\cdot\frac{S_{NO3}}{S_{NO3}+K_{NO3}^{HB}}\cdot X_{HB}$ |
| Anox_HB_NIR | 9 | $\mu_{NIR}^{HB}\cdot\eta_{HD}\cdot\frac{K_{I_O2_NIR}^{HB}}{S_{O2}+K_{I_O2_NIR}^{HB}}\cdot\frac{K_{I_NO_NIR}^{HB}}{S_{NO}+K_{I_NO_NIR}^{HB}}\cdot\frac{S_{S}}{S_{S}+K_{S_NIR}^{HB}}\cdot\frac{S_{NH4}}{S_{NH4}+K_{NH4}^{HB}}\cdot\frac{S_{NO2}}{S_{NO2}+K_{NO2}^{HB}}\cdot X_{HB}$ |
| Anox_HB_NOR | 10 | $\mu_{NOR}^{HB}\cdot\eta_{HD}\cdot\frac{K_{I_O2_NOR}^{HB}}{S_{O2}+K_{I_O2_NOR}^{HB}}\cdot\frac{S_{S}}{S_{S}+K_{S_NOR}^{HB}}\cdot\frac{S_{NH4}}{S_{NH4}+K_{NH4}^{HB}}\cdot\frac{S_{NO}}{S_{NO}+K_{NO}^{HB}+S_{NO}{}^{2}/K_{I_NO_NOR}^{HB}}\cdot X_{HB}$ |
| Anox_HB_NOS | 11 | $\mu_{NOS}^{HB}\cdot f(pH)\cdot\eta_{HD}\cdot\frac{K_{I_O2_NOS}^{HB}}{S_{O2}+K_{I_O2_NOS}^{HB}}\cdot\frac{K_{I_NO_NOS}^{HB}}{S_{NO}+K_{I_NO_NOS}^{HB}}\cdot\frac{S_{S}}{S_{S}+K_{S_NOS}^{HB}}\cdot\frac{S_{NH4}}{S_{NH4}+K_{NH4}^{HB}}\cdot\frac{S_{N2O}}{S_{N2O}+K_{N2O}^{HB}}\cdot X_{HB}$ |
| ***Lysis*** | | |
| AOB | 12 | $b_{AOB}\cdot\left(\frac{S_{O2}}{S_{O2}+K_{O2_b}}+\eta_{b,anox}\cdot\frac{K_{O2_b}}{K_{O2_b}+S_{O2}}\cdot\frac{S_{NOx}}{K_{NOx}+S_{NOx}}+\eta_{b,anaer}\cdot\frac{K_{O2_b}}{K_{O2_b}+S_{O2}}\cdot\frac{K_{NOx}}{K_{NOx}+S_{NOx}}\right)\cdot X_{AOB}$ |
| NOB | 13 | $b_{NOB}\cdot(\ldots)\cdot X_{NOB}$ |
| HB | 14 | $b_{HB}\cdot(\ldots)\cdot X_{HB}$ |
| ***Hydrolysis*** | | |
| Aerobic | 15 | $k_{H}\cdot\frac{X_{S}/X_{BH}}{K_{X}+X_{S}/X_{BH}}\cdot\frac{S_{O2}}{K_{O2}^{HB}+S_{O2}}\cdot X_{HB}$ |
| Anoxic | 16 | $k_{H}\cdot\eta_{ANOX}\cdot\frac{X_{S}/X_{BH}}{K_{X}+X_{S}/X_{BH}}\cdot\frac{K_{O2}^{HB}}{K_{O2}^{HB}+S_{O2}}\cdot\frac{S_{NOx-}}{K_{NO3}^{HB}+S_{NOx-}}\cdot X_{HB}$ |
| Anaerobic | 17 | $k_{H}\cdot\eta_{AN}\cdot\frac{X_{S}/X_{BH}}{K_{X}+X_{S}/X_{BH}}\cdot\frac{K_{O2}^{HB}}{K_{O2}^{HB}+S_{O2}}\cdot\frac{K_{NO3}^{HB}}{K_{NO3}^{HB}+S_{NOx-}}\cdot X_{HB}$ |

**Table S-I -2 –** Stoichiometric and kinetic parameter values. Initial values (literature) and estimated (corrected 20 C).

| Parameter | Definition | Value | Estimated | CV (%) | Units | Ref. |
|---|---|---|---|---|---|---|
| **AOB** | | | | | | |
| $K_{AOB.NH2OH}$ | S_NH2OH affinity for AOB | 0.2 | | | mgN/L | (1) |
| $K_{AOB.NH2OH.ND}$ | S_NH2OH affinity for AOB during NO reduction | 0.2 | | | mgN/L | (1) |
| $K_{AOB.NH3}$ | S_NH3 affinity for AOB | 0.0075 | 0.007 | 16.7 | mgN/L | (2) |
| $K_{AOB.NO.ND}$ | S_NO affinity for AOB | 0.015 | | | mgN/L | (1) |
| $K_{AOB.HNO2}$ | S_HNO2 affinity for AOB | 0.05 | | | µgN/L | (3) |
| $K_{AOB.O2.AMO}$ | S_O2 AMO-mediated affinity constant | 0.4 | | | mgO2/L | (2) |
| $K_{AOB.O2.HAO}$ | S_O2 HAO-mediated affinity constant | 0.4 | | | mgO2/L | (2) |
| $K_{AOB.O2.i}$ | S_O2 inhibition constant for AOB | 0.1 | | | mgO2/L | (4) |
| $K_{AOB.i.NH3}$ | S_NH3 inhibition constant for AOB | 10 | | | mgN/L | (5) |
| $K_{AOB.i.HNO2}$ | S_HNO2 inhibition constant for AOB | 0.75 | | | mgN/L | (5) |
| $\varepsilon_{AOB}$ | Reduction factor for HAO-mediated maximum reaction ra | 0.001 | 0.0031 | 3.2 | ( - ) | New |
| $\eta_{NIR}$ | Anoxic reduction factor for NO2 reduction | 0.15 | 0.22 | 4.5 | ( - ) | (5) |
| $\eta_{NOR}$ | Reduction factor for NO reduction | 0.15 | 0.36 | 5.6 | ( - ) | (5) |
| $\mu_{AOB.AMO}$ | Maximum AMO-mediated reaction rate | 0.78 | 0.49 | 2.3 | 1/d | (2) |
| $\mu_{AOB.HAO}$ | Maximum HAO-mediated reaction rate | 0.78 | | | 1/d | (2) |
| $b_{AOB}$ | Endogenous decay rate for AOB | 0.096 | | | 1/d | (2) |
| $Y_{AOB}$ | Yield coefficient for AOB | 0.18 | | | mgCOD/mgN | (2) |
| **NOB** | | | | | | |
| $K_{NOB.HNO2}$ | S_HNO2 affinity for NOB | 0.1 | 0.027 | 22.2 | µgN/L | (2) |
| $K_{NOB.O2}$ | S_O2 affinity constant for NOB | 0.5 | | | mgO2/L | (2) |
| $K_{NOB.i.NH3}$ | S_NH3 inhibition constant for NOB | 0.5 | | | mgN/L | (5) |
| $K_{NOB.i.HNO2}$ | S_HNO2 inhibition constant for NOB | 0.1 | | | mgN/L | (5) |
| $\mu_{NOB}$ | Maximum NOB growth | 0.78 | 1.04 | 10.6 | 1/d | (2) |
| $b_{NOB}$ | Endogenous decay rate for NOB | 0.096 | | | 1/d | (2) |
| $Y_{NOB}$ | Yield coefficient for NOB | 0.06 | | | mgCOD/mgN | (2) |
| **Others** | | | | | | |
| $f_{XI}$ | Fraction of inerts in biomass | 0.08 | | | ( - ) | (2) |
| $i_{NXB}$ | Nitrogen content of biomass | 0.086 | | | mgN/mgCOD | (2) |
| $i_{NXI}$ | Nitrogen content of inerts | 0.02 | | | mgN/mgCOD | (2) |
| $i_{NXS}$ | Nitrogen content of particulate | 0.06 | | | mgN/mgCOD | (2) |
| $\eta_{b,anox}$ | Anoxic reduction factor of endogenous decay | 0.7 | | | ( - ) | (6) |
| $\eta_{b,anaer}$ | Anaerobic reduction factor of endogenous decay | 0.33 | | | ( - ) | (6) |
| $K_{O2.b}$ | S_O2 affinity constant of endogenous decay | 0.2 | | | mgO2/L | (6) |
| $K_{NOx}$ | S_NO2+S_NO3 affinity constant of endogenous decay | 0.2 | | | mgN/L | (6) |
| $k_H$ | Hydrolysis rate | 2.21 | | | 1/d | (2) |
| $K_X$ | Affinity constant for hydrolysis | 0.15 | | | mgCOD/mgCOI | (2) |
| $\eta_{h,anox}$ | Anoxic hydrolysis factor | 0.7 | | | ( - ) | (6) |
| $\eta_{h,anaer}$ | Anaerobic hydrolysis factor | 0.33 | | | ( - ) | (6) |

| Parameter | Definition | Value | Estimated | CV (%) | Units | Ref. |
|---|---|---|---|---|---|---|
| **HB** | | | | | | |
| $K_{HB.NH4}$ | S_NH4 affinity constant for HB | 0.01 | | | mgN/L | (2) |
| $K_{HB.NO3}$ | S_NO3 affinity constant for HB | 0.2 | | | mgN/L | (6) |
| $K_{HB.NO2}$ | S_NO2 affinity constant for HB | 0.8 | | | mgN/L | (6) |
| $K_{HB.NO}$ | S_NO affinity constant for HB | 0.015 | | | mgN/L | (1) |
| $K_{HB.N2O}$ | S_N2O affinity constant for HB | 0.005 | 0.078 | 25.6 | mgN/L | (6) |
| $K_{HB.S}$ | S_S affinity constant for HB | 4.0 | | | mgCOD/L | (6) |
| $K_{HB.S.NAR}$ | S_S affinity constant for S_NO3 reduction | 5.0 | | | mgCOD/L | (6) |
| $K_{HB.S.NIR}$ | S_S affinity constant for S_NO2 reduction | 1.5 | 4.3 | 16.0 | mgCOD/L | (6) |
| $K_{HB.S.NOR}$ | S_S affinity constant for S_NO reduction | 2.4 | 5.3 | 15.7 | mgCOD/L | (6) |
| $K_{HB.S.NOS}$ | S_S affinity constant for S_N2O reduction | 2.0 | 4.1 | 9.8 | mgCOD/L | (6) |
| $K_{HB.O2}$ | S_O2 affinity constant for HB | 0.1 | | | mgO2/L | (2) |
| $K_{HB.O2.i.NAR}$ | S_O2 inhibition constant for S_NO3 reduction | 0.087 | | | mgO2/L | (6) |
| $K_{HB.O2.i.NIR}$ | S_O2 inhibition constant for S_NO2 reduction | 0.1 | | | mgO2/L | (6) |
| $K_{HB.O2.i.NOR}$ | S_O2 inhibition constant for S_NO reduction | 0.067 | | | mgO2/L | (6) |
| $K_{HB.O2.i.NOS}$ | S_O2 inhibition constant for S_N2O reduction | 0.031 | | | mgO2/L | (6) |
| $K_{HB.NO.i.NIR}$ | S_NO inhibition constant for S_NO2 reduction | 0.5 | | | mgN/L | (6) |
| $K_{HB.NO.i.NOR}$ | S_NO inhibition constant for S_NO reduction | 0.3 | | | mgN/L | (6) |
| $K_{HB.NO.i.NOS}$ | S_NO inhibition constant for S_N2O reduction | 0.075 | | | mgN/L | (6) |
| $\mu_{HB}$ | Maximum HB growth rate | 6.24 | 5.15 | 2.2 | 1/d | (6) |
| $\mu_{HB.NAR}$ | Maximum NO3-reduction reaction rate | 1.75 | 1.22 | 6.4 | 1/d | (6) |
| $\mu_{HB.NIR}$ | Maximum NO2-reduction reaction rate | 1 | 0.79 | 1.7 | 1/d | (6) |
| $\mu_{HB.NOR}$ | Maximum NO-reduction reaction rate | 2.18 | | | 1/d | (6) |
| $\mu_{HB.NOS}$ | Maximum N2O-reduction reaction rate | 2.18 | 0.83 | | 1/d | (6) |
| $\eta_{HD}$ | Reduction factor for HB denitrification | 0.2 | 0.33 | | ( - ) | (1) |
| $pH_{opt.nosZ}$ | Optimum pH for N2O-reduction | ( - ) | 7.9 | 1.3 | ( - ) | (7) |
| $w_{nosZ}$ | Sinusoidal parameter for N2O-reduction | ( - ) | 2.2 | 9.1 | ( - ) | (7) |
| $b_{HB}$ | Endogenous decay rate for HB | 0.41 | | | 1/d | (2) |
| $Y_{HB}$ | Yield coefficient for HB | 0.6 | | | mgCOD/mgCOD | (2) |

(1) Assumed this study, (2) Hiatt-Grady 2008, (3) von Schulthess 1994, (4) Ni 2011, (5) Park 2010, (6) Lisha Guo 2014

(7) Park 2007

| Parameter ↓ | \ Reference → | [1] | [2] | [3] | [4] | [5] | [6] | [6]* |
|---|---|---|---|---|---|---|---|---|
| $\varepsilon_{AOB}$, $\eta_{NN}$ | Reduction factor NN pathway | 0.0007 | 0.0015 | 0.12 | 0.001 | 0.026 | 0.0015 - 0.077 | 0.0007 - 0.0013 |
| $\eta_{ND}$ | Reduction factor ND pathway | 0.134 | 0.25 | 0.48 | 0.18 | 0.33 | 0.03 - 0.25 | 0.13 - 0.24 |

[1] Ni et al. 2014, [2] Pocquet et al. 2016, [3] Ding et al. 2016, [4] Ni et al. 2015, [5] Peng et al. 2017, [6] Lang et al. 2017 (model [2]), [6]* Lang et al. 2017 (model [1])

## S-II Sensitivity analysis on case study

Simulated conditions for the case study:

- Biomass composition: $f_{AOB}$ = 4.1%, $f_{NOB}$ = 1.8%, $f_{HB}$ = 74.1%, $f_{active}$ = 55%, VSS = 2.8 g/L.
- Reactor description: Vol = 3 L, $k_La_{DO}$ = 0.14 $min^{-1}$, $k_La_{NO}$ = $0.93k_La_{DO}$, $k_La_{N2O}$ = $0.89k_La_{DO}$, pH = 7.2, T = 22C.
- Oxygen concentrations for each condition: $DO_{Low}$ = 0.60 ± 0.20 mg/L, $DO_{High}$ = 2.10 ± 0.46 mg/L.

**Table S-II - 1** – Default uncertainty definition for model parameters.

| Parameter ($\theta_i$) | Uncertainty Class |
|---|---|
| $i_{NXB}$, $i_{NXI}$, $i_{NXS}$, VSS, $Y_{AOB}$, $Y_{HB}$, $Y_{NOB}$, $K_La_{N2O}$, $K_La_{O2}$, $K_La_{NO}$, $f_{active}$, $f_{AOB}$, $f_{NOB}$, $f_{HB}$, $f_I$, | 1 (± 10%) |
| $b_{AOB}$, $b_{HB}$, $b_{NOB}$, $k_H$, $\eta_{HD}$, $\mu_{AOB_AMO}$, $\mu_{AOB_HAO}$, $\mu_{HB}$, $\mu_{HB_NAR}$, $\mu_{HB_NIR}$, $\mu_{HB_NOR}$, $\mu_{HB_NOS}$, $\mu_{NOB}$, $\eta_{b_anox}$, $\eta_{b_anaer}$, $\eta_{h_anox}$, $\eta_{h_anaer}$, | 2 (± 25%) |
| $K_{AOB_NH2OH}$, $K_{AOB_NH2OH_ND}$, $K_{AOB_NH3}$, $K_{AOB_NO_ND}$, $K_{AOB_HNO2}$, $K_{AOB_O2_AMO}$, $K_{AOB_O2_HAO}$, $K_{AOB_O2_i}$, $K_{HB_NH4}$, $K_{HB_N2O}$, $K_{HB_NO}$, $K_{HB_NO2}$, $K_{HB_NO3}$, $K_{HB_O2}$, $K_{HB_O2_i_NAR}$, $K_{HB_O2_i_NIR}$, $K_{HB_O2_i_NOR}$, $K_{HB_O2_i_NOS}$, $K_{HB_S}$, $K_{HB_S_NAR}$, $K_{HB_S_NIR}$, $K_{HB_S_NOR}$, $K_{HB_S_NOS}$, $K_{NOB_HNO2}$, $K_{NOB_O2}$, $K_{NOB_i_HNO2}$, $K_{NOB_i_NH3}$, $K_{AOB_i_HNO2}$, $K_{AOB_i_NH3}$, $K_{b_O2}$, $K_X$, $\varepsilon_{AOB}$, $\eta_{NIR}$, $\eta_{NOR}$, | 3 (± 50%) |

**Table S-II - 2** – Standardized regression coefficients ($\beta$) for NO and $N_2O$ during $NH_4^+$ oxidation.

| | Low DO (≈ 0.5 mg/L) | | | | High DO (≈ 2.0 mg/L) | | | |
|---|---|---|---|---|---|---|---|---|
| | $N_2O$ | | NO | | $N_2O$ | | NO | |
| Rank | $\theta_i$ | $\beta_i$ | $\theta_i$ | $\beta_i$ | $\theta_i$ | $\beta_i$ | $\theta_i$ | $\beta_i$ |
| 1. | $\mu_{NOB}$ | -0.29 | $\eta_{NOR}$ | -0.24 | $\mu_{AOB.HAO}$ | 0.35 | $\eta_{NOR}$ | -0.33 |
| 2. | $K_{HB.O2.i.NIR}$ | 0.27 | $K_{AOB.O2.i.NIR}$ | 0.21 | $\mu_{AOB.AMO}$ | 0.33 | $K_{AOB.NO.ND}$ | 0.25 |
| 3. | $\mu_{AOB.AMO}$ | 0.27 | $K_{AOB.O2.HAO}$ | -0.20 | $\mu_{NOB}$ | -0.29 | $\mu_{NOB}$ | -0.25 |
| 4. | $K_{NOB.O2}$ | 0.26 | $K_{NOB.O2}$ | 0.20 | $K_{HB.O2.i.NIR}$ | 0.27 | $\mu_{AOB.HAO}$ | 0.21 |
| 5. | $K_{HB.O2.i.NOS}$ | -0.23 | $\mu_{NOB}$ | -0.19 | $K_{HB.S.NIR}$ | -0.22 | $K_{HB.S.NIR}$ | -0.21 |
| 6. | $K_{AOB.O2.AMO}$ | -0.22 | $\mu_{AOB.AMO}$ | 0.17 | $K_{NOB.HNO2}$ | 0.21 | $\mu_{AOB.AMO}$ | 0.20 |
| 7. | $\mu_{AOB.HAO}$ | 0.22 | $K_{HB.O2.i.NOR}$ | -0.17 | VSS | 0.17 | $K_{HB.O2.i.NIR}$ | 0.20 |
| 8. | $K_{HB.S.NIR}$ | -0.22 | $K_{HB.S.NIR}$ | -0.15 | $K_{HB.O2.i.NOS}$ | -0.16 | $K_{NOB.HNO2}$ | 0.16 |
| 9. | $K_{HB.S.NOS}$ | 0.19 | $K_{AOB.NO.ND}$ | 0.14 | $K_{HB.NO2}$ | -0.16 | $K_{HB.NO2}$ | -0.15 |
| 10. | $K_{AOB.O2.HAO}$ | -0.18 | $K_{HB.S.NOR}$ | 0.12 | $\mu_{HB.NIR}$ | 0.15 | $\mu_{HB.NIR}$ | 0.13 |

## S-III Information content of Experimental Design

Results from the sensitivity analysis of experiments: B ($N_2O$ and NO) and C (DO).

**Table S-III - 1** – Top ranked parameters of the GSA during NO and $N_2O$ production of experiments (ii)

| | **Ranking ($\beta^2$)** $N_2O$ | **NO** |
|---|---|---|
| 1 | $K_{HB.N2O}$ | $K_{HB.NO}$ |
| 2 | $K_{HB.NO2}$ | $K_{HB.NO2}$ |
| 3 | $\mu_{HB.NOS}$ | $\mu_{HB.NOR}$ |
| 4 | $K_{HB.S.NOS}$ | $K_{HB.S.NOR}$ |
| 5 | $\mu_{HB.NIR}$ | $\mu_{HB.NIR}$ |
| 6 | $K_{HB.S.NIR}$ | $Y_{HB}$ |
| 7 | $\eta_{HD}$ | $K_{HB.S.NIR}$ |
| 8 | $Y_{HB}$ | $\eta_{HD}$ |

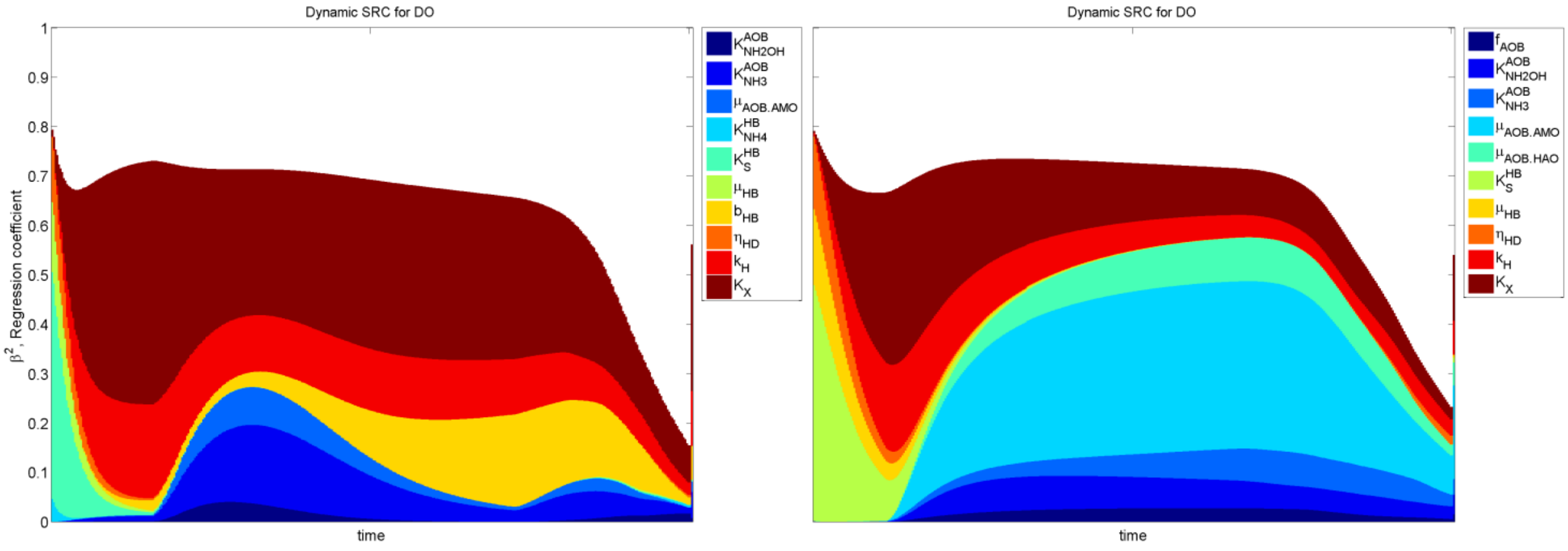

**Figure S-III – 1** - Dynamic GSA for DO in experiments (C) after $NH_4^+$ pulses (left, 2 low $NH_4^+$; right, 1 high $NH_4^+$).

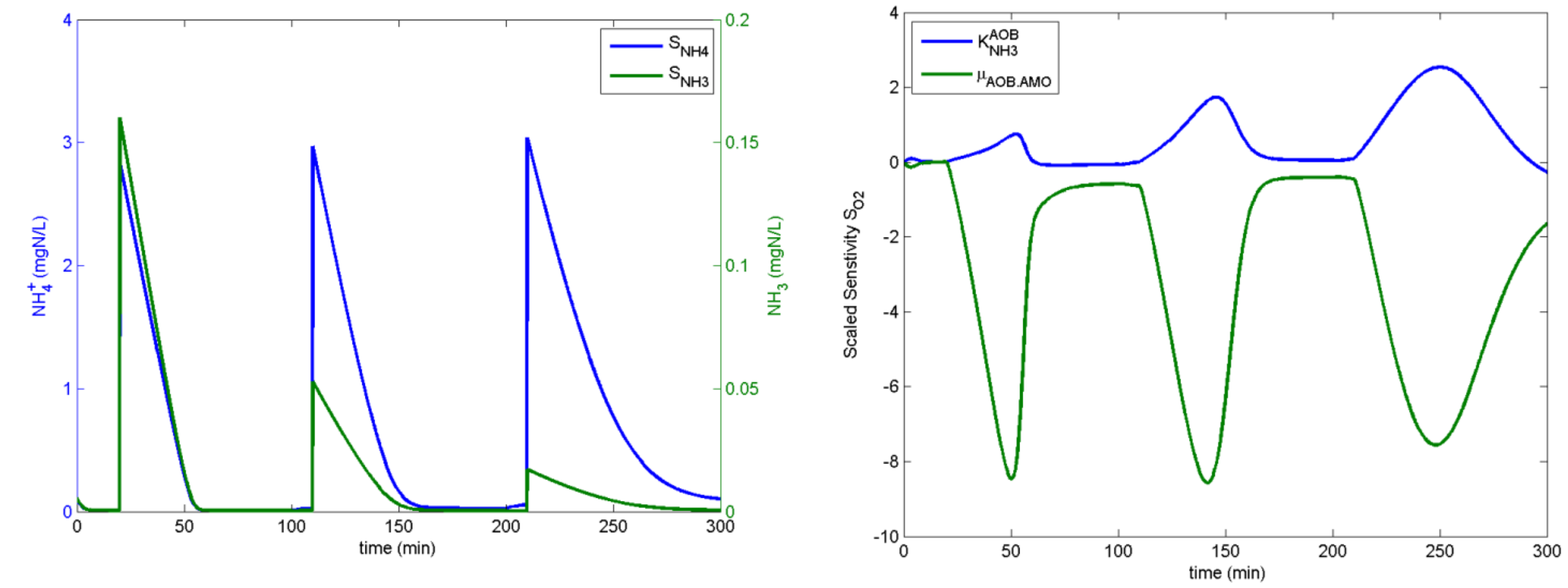

**Figure S-III – 2** – Local scaled sensitivity function (Brun et al., 2001) for DO after three equal $NH_4^+$-N pulses at excess DO at, from left to right, pH = 8, 7.5 and 7.

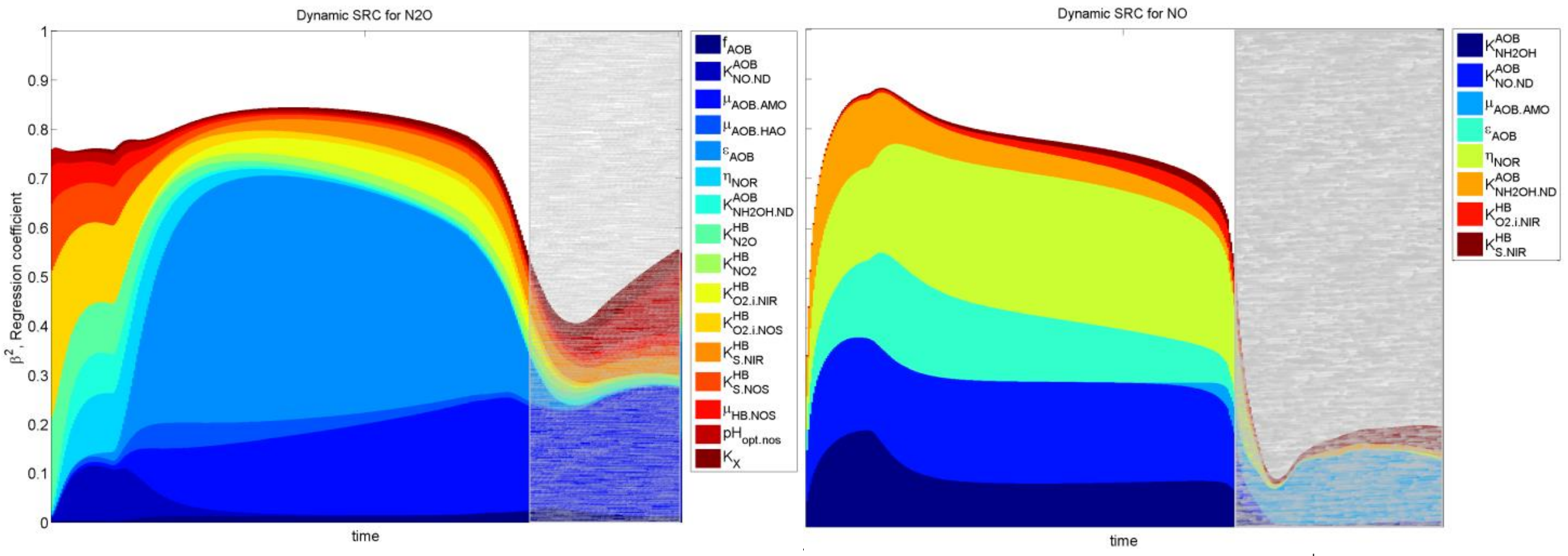

**Figure S-III – 3** - Dynamic GSA for $N_2O$ (left) and NO (right) for one experiment (C) after $NH_4^+$ pulse. The shaded area corresponds to $NH_4^+ \approx 0$ mgN/L where no reaction occurs.

## S-IV → Parameter estimation results

**Table S_IV - 1** – Best-fit simulation analysis of residuals.

| **Variable →** | **$NH_4^+$** | | **$NO_3^-$** | | **$NO_2^-$** | | **$N_2O$** | | **NO** | | **DO** | |
|---|---|---|---|---|---|---|---|---|---|---|---|---|
| **Scenario ↓** | **$R^2$** | **F-test** | **$R^2$** | **F-test** | **$R^2$** | **F-test** | **$R^2$** | **F-test** | **$R^2$** | **F-test** | **$R^2$** | **F-test** |
| (A) Hydrolysis | 0.974 | 1 | | | | | | | | | | |
| (A) NAR-NIR-NOS | | | 0.991 | 1 | 0.966 | 1 | 0.980 | 1 | | | | |
| (B) K_HB_S | | | | | | | 0.936 | 1 | 0.937 | 1 | | |
| (C) HB aerobic | | | | | | | | | | | 0.998 | 1 |
| (C) NOB | | | | | | | | | | | 1.000 | 1 |
| (C) AOB | | | | | | | | | | | 0.999 | 0 |
| (C) $N_2O$ | | | | | | | 0.994 | 1 | 0.987 | 1 | | |

**Table S-IV - 2** – Correlation matrices of the estimated parameters.

| | | | | | | |
|---|---|---|---|---|---|---|
| $\mu_{HB_NAR}$ | 1 | | | | | |
| $\mu_{HB_NIR}$ | 0.6 | 1 | | | | |
| $\mu_{HB_NOS}$ | | | 1 | | | |
| $K_{HB_S_NIR}$ | | | | 1 | | |
| $K_{HB_S_NOR}$ | | | | 0.4 | 1 | |
| $K_{HB_S_NOS}$ | | | | 0.25 | -0.03 | 1 |

| | | | | | |
|---|---|---|---|---|---|
| $K_{AOB_NH3}$ | 1 | | | | |
| $\mu_{AOB_AMO}$ | 0.65 | 1 | | | |
| $\mu_{HB}$ | | | 1 | | |
| $K_{NOB_HNO2}$ | | | | 1 | |
| $\mu_{NOB}$ | | | | 0.95 | 1 |

| | | | |
|---|---|---|---|
| $\eta_{NIR}$ | 1 | | |
| $\varepsilon_{AOB}$ | -0.16 | 1 | |
| $\eta_{NOR}$ | 0.15 | 0.71 | 1 |

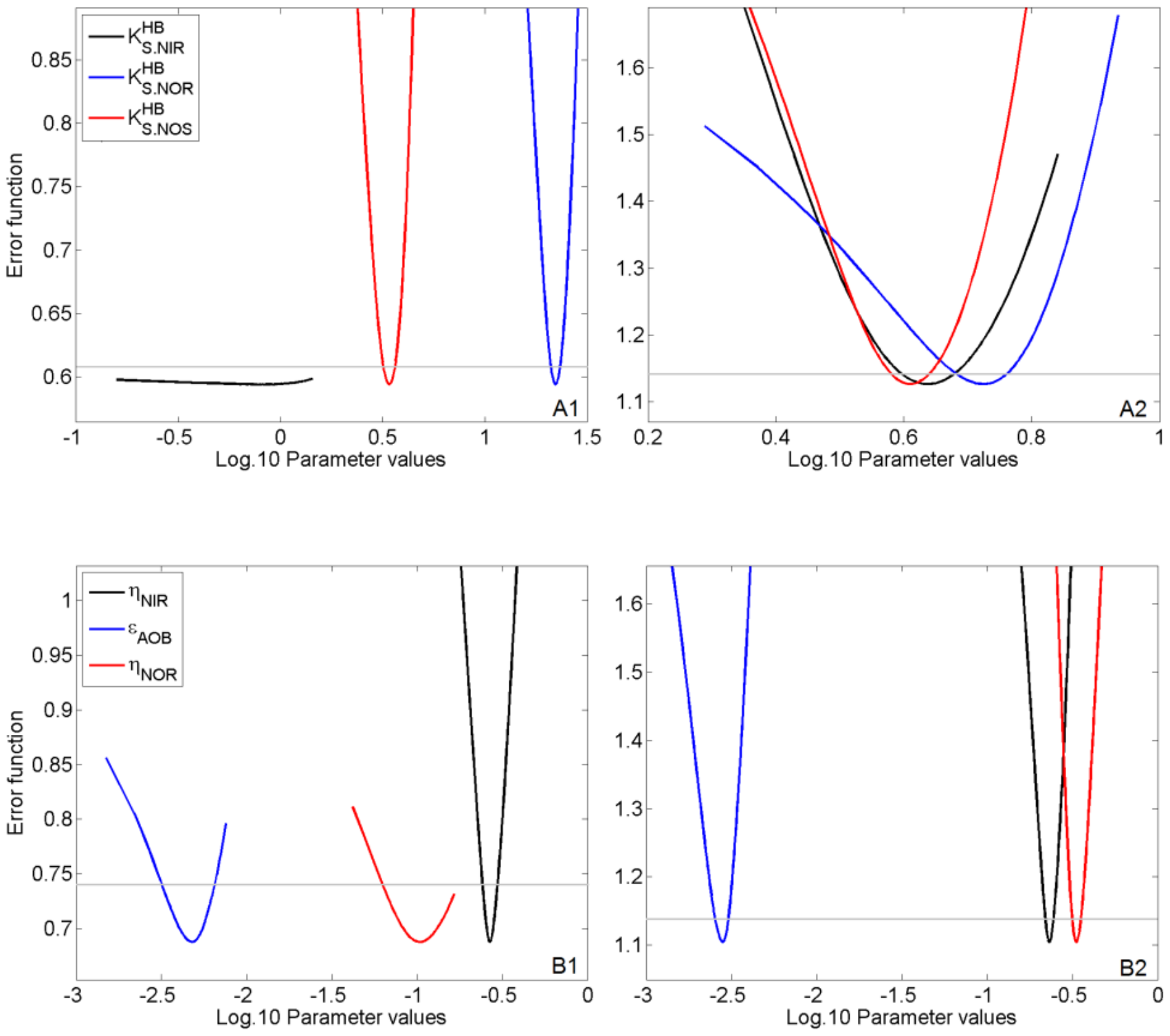

**Figure S-IV - 1** – Error function profile for experiments: B, (top) and C (bottom). Error from $N_2O$ datasets (left), Error from NO + $N_2O$ datasets (right). Parameters were varied one at a time. Grey line represents the 95% confidence interval for individual parameters.

## Analysis of residuals

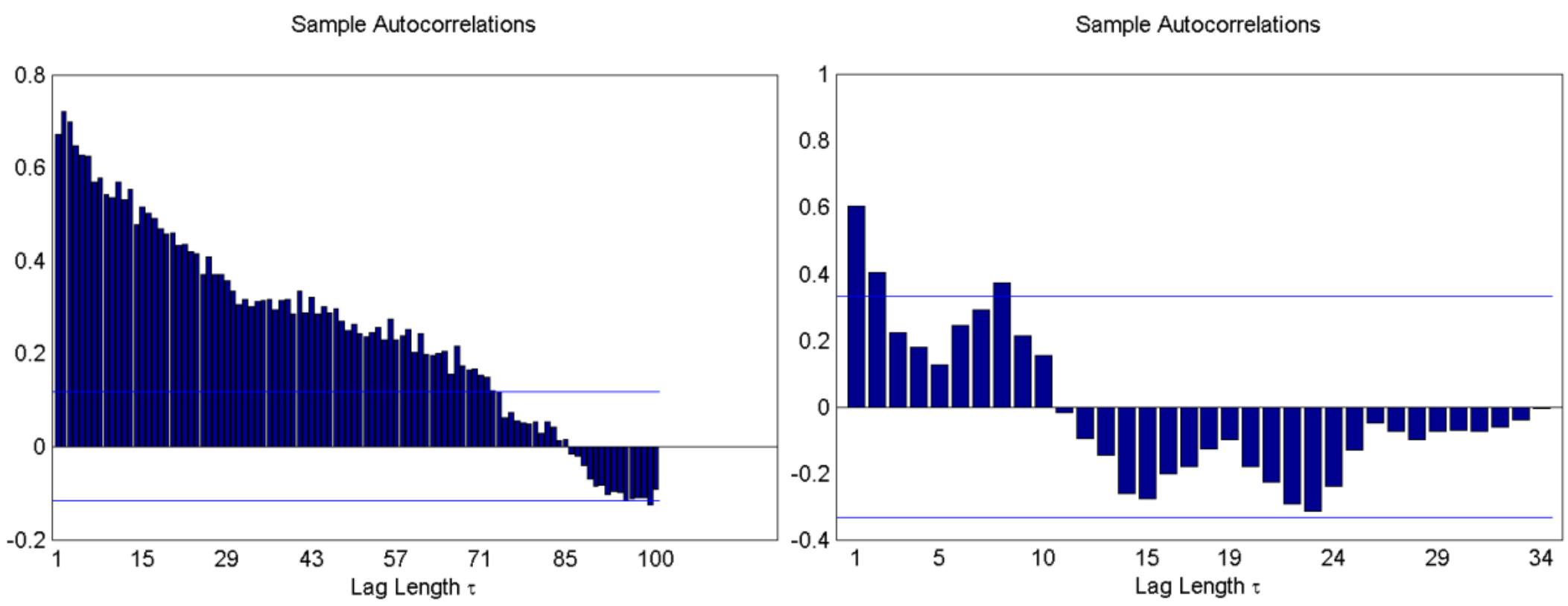

**Figure S-IV – 2** – Example of autocorrelation of residuals for one experiment (C): original DO dataset (left, 280 points), DO dataset after down-sampling from 1point/30 seconds to 1point/240 seconds (35 points) (right).

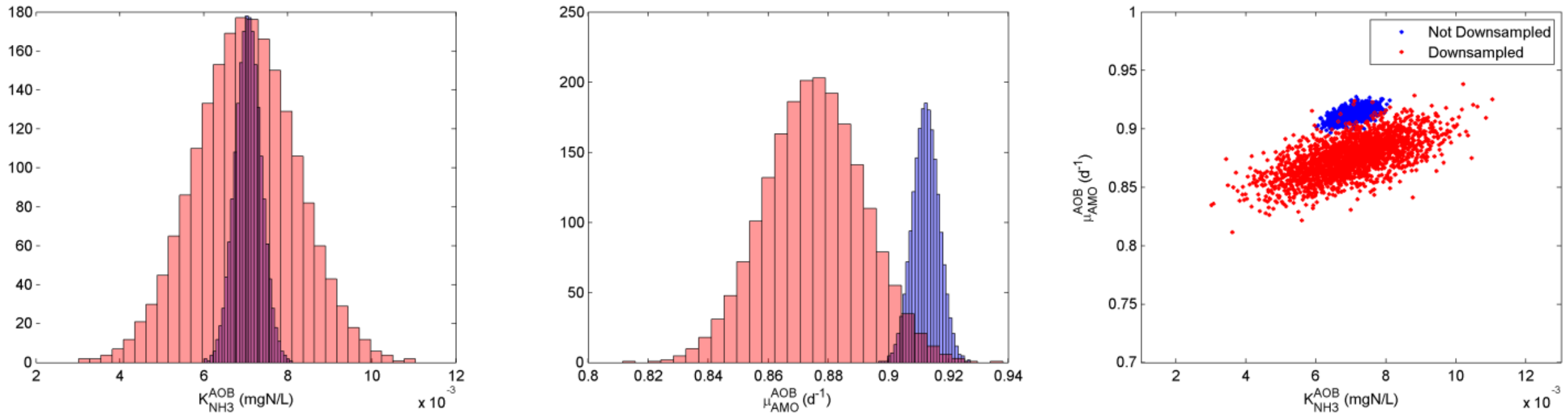

**Figure S-IV – 3** – Histogram and scatter plot of $K_{AOB.NH3}$ and $\mu_{AOB.AMO}$ best-fit estimates for the original DO dataset (blue, CV = 4.4 – 0.5% respectively) and down-sampled dataset (red, CV = 16.7 – 1.8%). 2000 samples were randomly selected from a bivariate normal distribution. The down-sample frequency was, in average, from 1point/30 seconds to 1point/ 815 seconds.

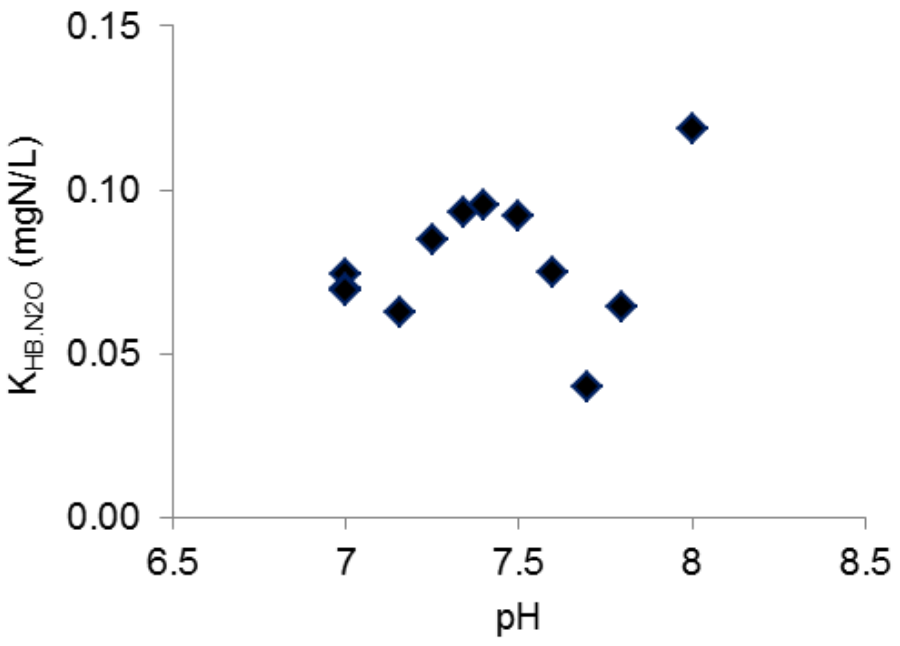

**Figure S-IV - 4** – $K_{HB.N2O}$ best-fit estimates from experiments (A) at varying pH ($R^2$ = 0.062, n=12)

## S-V Experimental Results

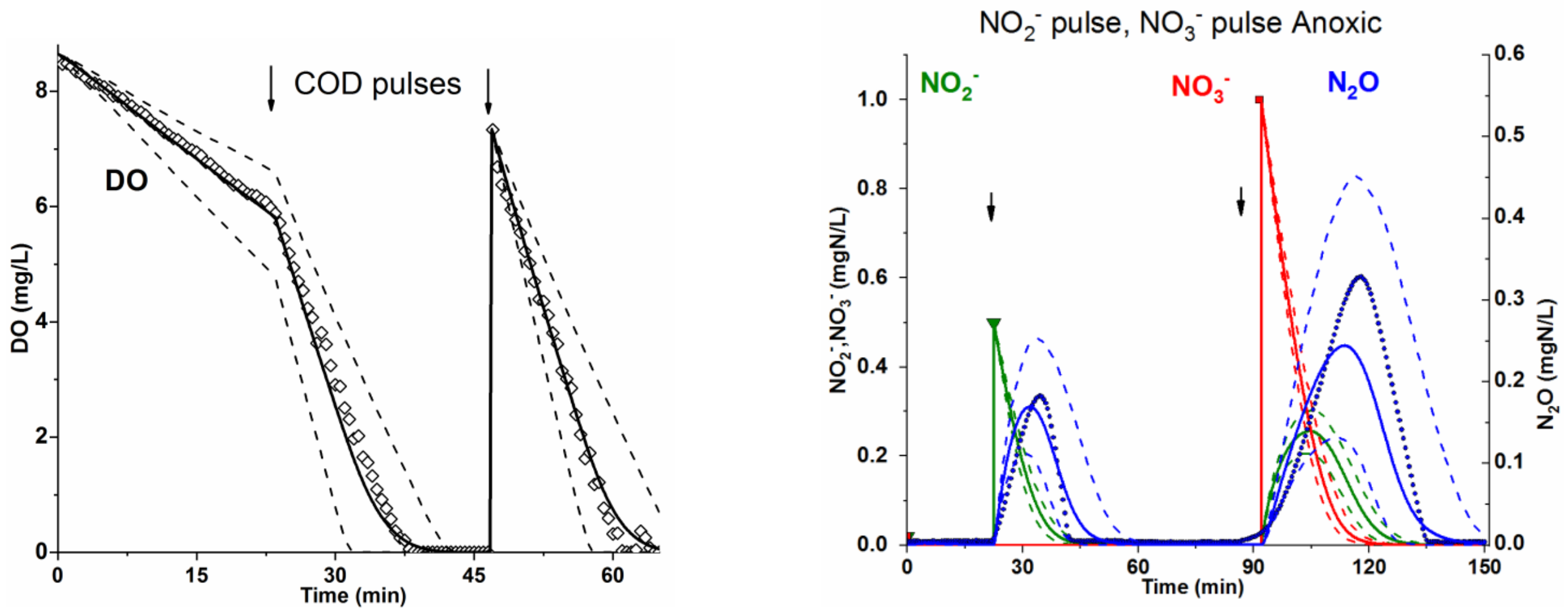

**Figure S-V – 1** - Experimental and modelling results obtained during Scenario A (left, aerobic COD removal) and B (right, anoxic $NO_2^-$ and $NO_3^-$ reduction) after parameter estimation. Experimental data (markers), best-fit simulations (solid lines), 95% confidence intervals for the uncertainty of all model parameters (dashed lines).

**Table S-V - 1** – Maximum process rates for heterotrophic denitrification and aerobic growth.

| | | Literature | | This study | | |
|---|---|---|---|---|---|---|
| **Process** | **Units** | **Endogenous** | **Excess COD** | **Endogenous** | **Excess COD** | **References** |
| **$NO_3^-$ reduction** | (mgN/gVSS.h) | 0.8 - 4.9 | 2.4 - 28 | 1.3 - 1.7 | 5.7 - 8.3 | (1), (2), (3), (4) |
| **$NO_2^-$ reduction** | (mgN/gVSS.h) | 1.7 - 3.3 | 12 - 22.5 | 2.5 - 2.6 | 6.2 | (4), (5) |
| **$N_2O$ reduction** | (mgN/gVSS.h) | 2.4 - 3.0 | 18 - 52 | 4.7 | 10 - 14 | (4) |
| **Heterotrophic OUR** | (mgCOD/gVSS.h) | 2 - 16 | 60 | 4.5 - 7 | 34.6 | (3), (6) |

(1) Schulthess et al. 1995, (2) Ni et al. 2008, (3) Li and Ju 2002, (4) Yang et al. 2009, (5) Ribera-Guardia et al. 2014, (6) Benes et al. 2002

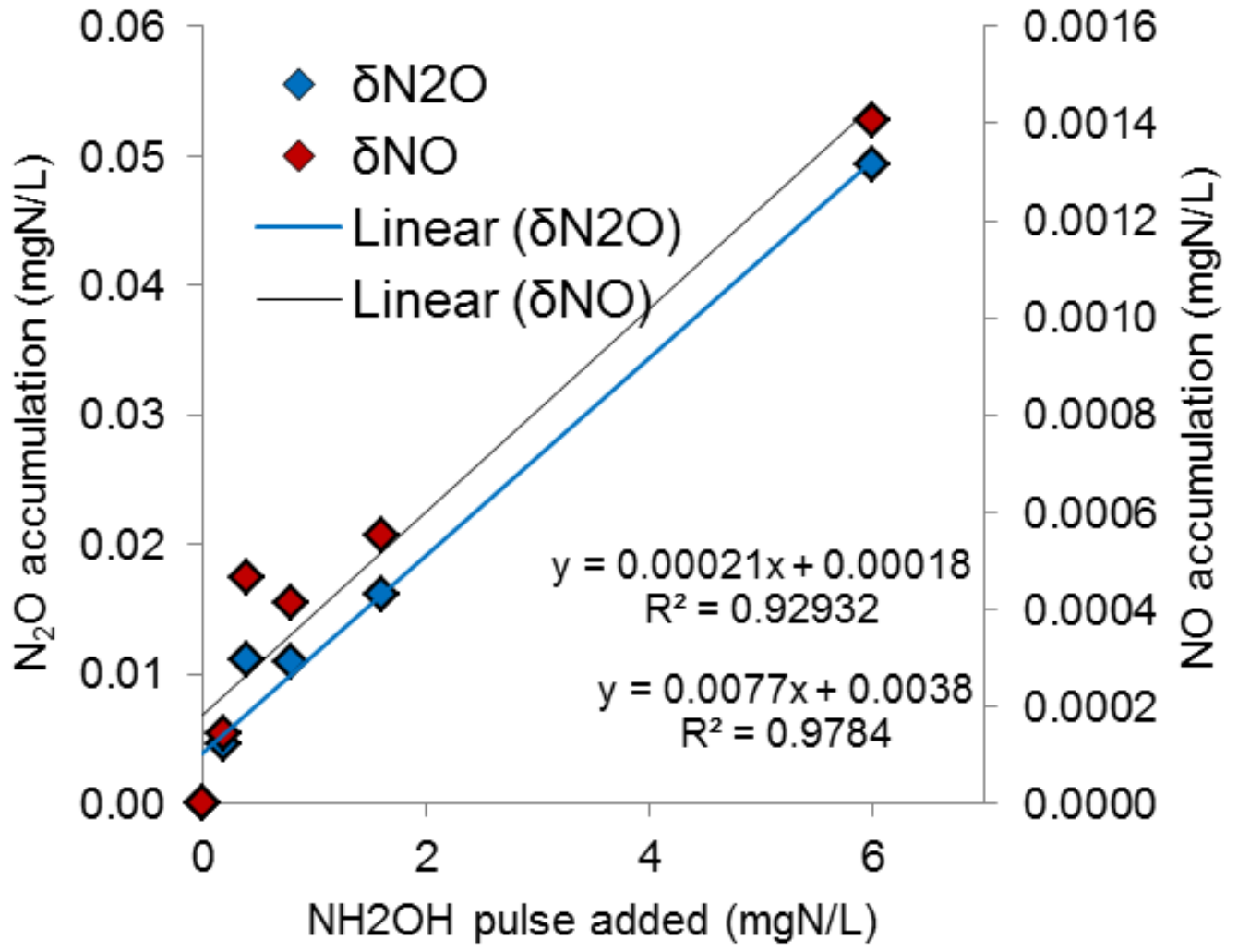

**Figure S-V - 2** – $N_2O$ and NO bulk accumulation for $NH_2OH$ pulses added to nitrogen-free mineral medium.

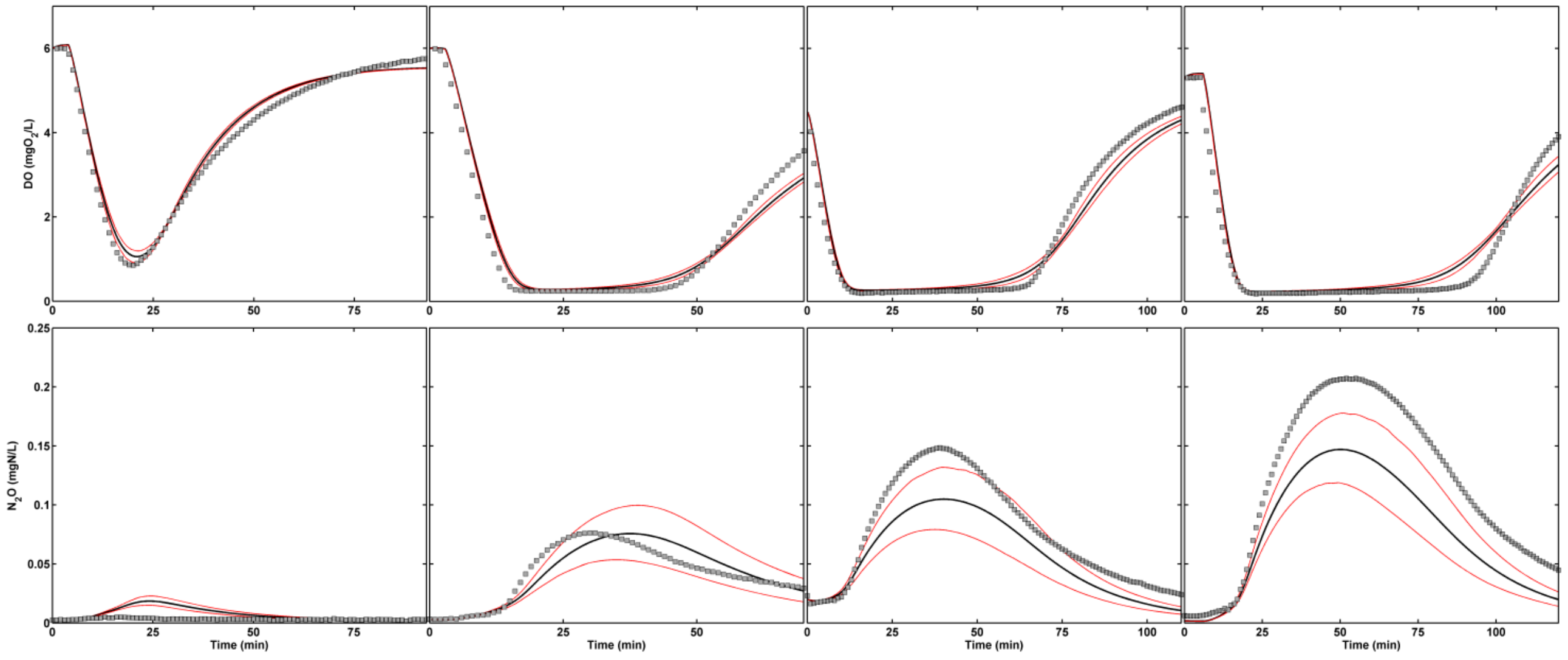

**Figure S-V - 2** – Model evaluation results (D): DO consumption and associated $N_2O$ production after $NH_4^+$ pulses (2.3, 3.6, 4.7, 5.5 mgN/L) at constant aeration.

## S-VI Uncertainty propagation

**Table S-VI - 1** – $NH_4^+$ removal, $N_2O$ emissions and pathway contribution for the estimated parameters. Uncertainty calculated by the Monte Carlo method (Sin et al., 2009), n = 500 samples.

| | | **DO = 0.5 mg/L** | **DO = 2.0 mg/L** |
|---|---|---|---|
| ΔTN (mgN/L) | | 16.8 ± 0.1 | 27.1 ± 0.3 |
| $N_2O_{emitted/removed}$ | | 4.6 ± 0.6% | 1.2 ± 0.1% |
| $N_2O_{pathway_contrib}$ | NN | 19 ± 2% | 51 ± 3% |
| | ND | 64 ± 2% | 42 ± 3% |
| | HD | 17 ± 2% | 7 ± 2% |